\documentclass[pdflatex,sn-mathphys-num]{sn-jnl}

\usepackage{amsmath}
\usepackage{amsfonts}
\usepackage{amssymb}
\usepackage{graphicx, rotating}
\usepackage{epstopdf}
\usepackage{epsfig}
\usepackage{latexsym}
\usepackage{multirow}
\usepackage{slashed}
\usepackage{amstext}
\usepackage{array}  
\usepackage{nicefrac}
\usepackage{bbold}
\usepackage{tikz}
\usetikzlibrary{shapes.geometric, arrows.meta, positioning}
\usepackage{xcolor}
\usepackage[normalem]{ulem}
\usepackage{xspace}

\usepackage[utf8]{inputenc}
\usepackage{bm}
\usepackage{leftidx}

\usepackage{makecell}

\usepackage{colortbl}

\newcommand{\be}{\begin{equation}}
\newcommand{\ee}{\end{equation}}
\newcommand{\ba}{\begin{array}}
\newcommand{\ea}{\end{array}}

\definecolor{LavGrey}{HTML}{93a1c5}
\definecolor{Fern}{HTML}{588157}
\definecolor{DuskBlue}{HTML}{274c77}
\definecolor{RosyCopper}{HTML}{dd614a}
\definecolor{PaperWhite}{HTML}{fdfeff}

\newcommand{\cnb}{C$\nu$B\xspace}

\begin{document}

\title{The KM3NeT event: a primordial high energy neutrino?}

\author*[1,2]{\fnm{Nicolas} \sur{Grimbaum Yamamoto}}\email{nicolas.grimbaumyamamoto@physics.ox.ac.uk}
\author[1]{\fnm{Thomas} \sur{Hambye}}\email{thomas.hambye@ulb.be}

\affil[1]{Service de Physique Th\'eorique, Universit\'e Libre de Bruxelles, Boulevard du Triomphe, CP225, 1050 Brussels, Belgium}
\affil[2]{Rudolf Peierls Centre for Theoretical Physics, University of Oxford, Parks Road, Oxford OX1 3PU, United Kingdom}

\abstract{
High energy neutrinos can be injected in the early Universe from the decay or annihilation of long lived primordial relics. We analyse the possibility that the ultrahigh energy neutrino event recently observed by the KM3NeT neutrino telescope could have such an origin. This possibility has the advantage of leading to a sharp spectral feature in such a way that the neutrino flux can be small at all energies except at the KM3NeT event energy. Thus, along this scenario the tension with null results from other experiments is reduced with respect to the usual power law case analysed by the KM3NeT and IceCube experiments, down to nearly the irreducible tension between these datasets. 
At such energies and for an emission around the recombination time, interactions of these neutrinos with background neutrinos prove to be relevant and must be determined from the development of a dedicated code. These interactions, as well as final state radiation processes, modify the spectrum.
Interestingly, it turns out that the scenario can also leave an imprint in the CMB that could be probed in the near future.
Interestingly too, this scenario does not predict an associated $\gamma$-ray flux beyond observation.
All in all we do find that the high energy neutrino could be a primordial high energy neutrino, provided it has been produced around the
recombination time or later. More generally, the KM3NeT event constitutes an excellent case study of how the spectrum of primordial high energy neutrinos is distorted by medium effects.}

\maketitle

\section{Introduction}

Last year, the KM3NeT neutrino telescope reported the observation of a neutrino with unprecedented high energy~\cite{KM3NeT:2025npi}. The  neutrino energy (from the observed muon energy $E_\mu=120^{+110}_{-60}$~PeV) that best fits the data is equal to 220 PeV, with a central 90\% range of 72~PeV--2.6~EeV. 
This observation is surprising in the light of the non-observation of events at similar energies by the IceCube~\cite{IceCube:2018fhm,IceCubeCollaborationSS:2025jbi} and Pierre Auger~\cite{PierreAuger:2023pjg} collaborations,
despite many years of data taking.
Combining the upper limits from these 2 observatories at the KM3NeT event energies with the sensitivity of KM3NeT, the tension is at the level of $2.01\sigma$.
In other words the observation of the KM3NeT event would be due to an overfluctuation, with probability at the level of $\sim 1/45$.
Assuming in addition a power law to account for the observation of neutrino events by IceCube at lower energies (the $\sim \text{PeV}$ HESE selection~\cite{IceCube:2020wum}), the tension rises to the level of $2.56\sigma$, corresponding to an overfluctuation with probability $\sim 1/190$.

In this work, we investigate the possibility that the KM3NeT neutrino would not be associated with a (e.g.~astrophysical) power law, but due to a (e.g.~particle physics induced) sharp spectral feature of the neutrino spectrum,
i.e. a sharp rise of the flux at energies not far below the one of the KM3NeT event and possibly also a sharp fall beyond. This can be induced by the very slow decay of super heavy dark matter (DM) particles, a possibility that has been studied in several analyses, see e.g.~\cite{Kohri:2025bsn,Borah:2025igh,Su:2025qzt, Barman:2025hoz}. For instance, a decay into a neutrino and a standard model (SM) particle leads to a monochromatic galactic neutrino flux as well as an extragalactic flux. 
The DM decay lifetime required is $\sim10^{29}$ seconds, see also the lower limit from galactic constraints that hold on the lifetime for such a channel from IceCube~\cite{IceCube:2023ies}.
For this kind of decay, some tension can nevertheless survive from the large $\gamma$-ray flux that the SM particles produce. A decay into two neutrinos can somewhat reduce this tension. Primordial black holes can also produce features in the neutrino spectrum by evaporating in the vicinity of Earth, but the associated flux of SM particles that are necessarily produced during the evaporation of the black hole puts stringent constraints on this scenario from the non-observation of a $\gamma$-ray flux~\cite{Airoldi:2025opo}.

In this work, we investigate the possibility of a sharp spectral feature that could result from the production of high energy neutrinos in the early Universe, which would have travelled all the way to the detector today. Such a primordial high energy neutrino (Phenu for short) could have been induced, in the simplest case, from the decay of a relic particle into two neutrinos or into a neutrino and another particle. As a result of the fact that not all decays occur at the same time, so that the produced neutrinos are not all redshifted in the same way, the primordial high energy neutrino spectrum is not monochromatic as at injection but can be still quite sharp today. 
Curiously, this possibility has been relatively little studied. Ref.~\cite{Frampton:1980rs}
has put forward the possibility of an early Universe source particle decaying into two neutrinos, basically without studying it, and Ref.~\cite{Gondolo:1991rn} considers it, in particular from the perspective of early experimental constraints, see also refs.~\cite{Kanzaki:2007pd,Ema:2013nda, Ema:2014ufa,McKeen:2018xyz,Berghaus:2018zso,Jaeckel:2020oet} in specific contexts, cases, and energy ranges. 
Recently, ref.~\cite{GrimbaumYamamoto:2025nai} has studied and determined in a systematic way the parameter space (i.e.~decay lifetime $\tau_P$, mass $m_P$ and abundance of the source particle $P$) in terms of theoretical constraints (mainly BBN and CMB constraints), experimental constraints, and possible interactions of these neutrinos with cosmic neutrinos or between themselves on their way to Earth. Too many interactions 
would smooth so much the energy spectrum that no spectral feature would remain.
A relatively large parameter space turns out to be allowed by these theoretical constraints without large alteration of the energy spectrum, with lifetime that can be as short as $10^{9}$ seconds (i.e.~emission well before recombination time $t_\text{rec}\sim 10^{13}$~sec), and masses from tens of GeV to $\sim 10^{11}$ GeV. Given current experimental sensitivities, such a Phenu could be observed today with any energy ranging from a fraction of GeV to tens or hundreds of PeV.

A $\sim 200$~PeV neutrino today implies a $10^{11}$~GeV mass for the source particle if the neutrino is emitted for example around the recombination time. As ref.~\cite{GrimbaumYamamoto:2025nai} shows, for such high energies one expects interactions with the cosmic neutrino background (C$\nu$B) to be quite relevant, whereas self-interactions of two Phenus are irrelevant. Thus, a proper determination of the distortion of the energy spectrum by interactions with the cosmic neutrinos must be performed. This requires the development of a dedicated Monte Carlo code. In this work we developed such a code, and use it to determine how sharp the spectrum remains as a function of the parameters for energies today relevant for the KM3NeT event. More generally, to develop such a code is also interesting because it could be further used to compute the neutrino energy spectrum at the detector in any scenario involving production of neutrinos in the early Universe. 
The KM3NeT event provides a concrete case study of how a primordial high energy neutrino spectrum can be distorted by interactions with the medium (the parameters reproducing this event lie precisely in a region where these effects are important without completely erasing the original spectral feature), an effect that could be relevant in many other contexts.
More generally, the other way around, the observation of a distorted neutrino spectrum, if due to primordial high energy neutrinos, would, through the code, provide an indirect signature of the C$\nu$B itself, and allow one to probe it.

A proper determination of the spectrum also requires the determination of the final state radiation (FSR) that results from radiative corrections at production. As is well known, these grow with energy and are expected to create numerous lower energy neutrinos as well as distort the high energy spectrum. 
At such high energies, we use the \texttt{HDMSpectra} package~\cite{Bauer:2020jay}, which extends up to the Planck scale, rather than \texttt{PYTHIA}-based tools~\cite{Bierlich:2022pfr} such as \texttt{CosmiXs}~\cite{Arina:2023eic}, which include electroweak corrections but are tabulated only up to source particle masses of 100~TeV.

A particularity of the Phenu scenario is that, in contradistinction to the DM decay scenario, the flux of any other high energy particle that could be produced in association with the neutrinos at the source, would be suppressed in the detector. Thus the scenario is not expected to be in tension with observations of, for instance, $\gamma$-rays. This is true even if the decay produces not only primary neutrinos but also other primary Standard Model particles, such as photons. One must nevertheless make sure that such an injection of high energy neutrinos in the early Universe is allowed by cosmological constraints, in particular CMB constraints.
Note, interestingly, that at such high energies the constraints on the injection of other particles (such as charged leptons) are hardly stronger than on the injection of neutrinos.\footnote{This is due to the fact that at such high energies the fraction of electromagnetic or hadronic energy that a neutrino or charged lepton injects differs from the total energy injected by a factor of ${\cal O}(1)$, see e.g.~\cite{Bianco:2025boy}.} Thus, for example, an associated production of charged lepton pairs at the same level as the neutrino pairs (as in many scenarios due to $SU(2)_L$ invariance) would be hardly more constrained than a pure neutrino injection scenario.\footnote{See the discussion of models that can lead to such a pure neutrino injection scenario in~\cite{Bianco:2025boy}. This also holds for 3-body decay scenarios with at least one neutrino in the final state, which as shown in~\cite{GrimbaumYamamoto:2025nai} also leads to a sharp spectral feature.}

We find that adding a Phenu component to the single power law describing the diffuse astrophysical spectrum reduces the tension from $2.56\sigma$ to about $2.1\sigma$.
This is close to the irreducible $2.0\sigma$ tension between the observation of the KM3NeT event and the non-observation of events by IceCube and Auger at the same energies.

This work is presented as follows. 
Section~\ref{sec:fsr} briefly discusses the spectrum at injection including FSR.
Section~\ref{sec:interactions} presents the Monte Carlo code that we developed to compute the distortion experienced by the spectrum due to interactions with the \cnb.
In section~\ref{sec:characterisation}, we characterise the resulting flux today, and in section~\ref{sec:fit} we apply these results to the case of the KM3NeT event and fit the existing data, taking into account IceCube and Auger data.
We then recall the existing cosmological constraints in section~\ref{sec:constraints}, and discuss $\gamma$-ray constraints in section~\ref{sec:checks}.
Finally, we conclude in section~\ref{sec:summary}.

\section{Final state radiation}
\label{sec:fsr}

\begin{figure}
    \centering
    \includegraphics[width=0.5\linewidth]{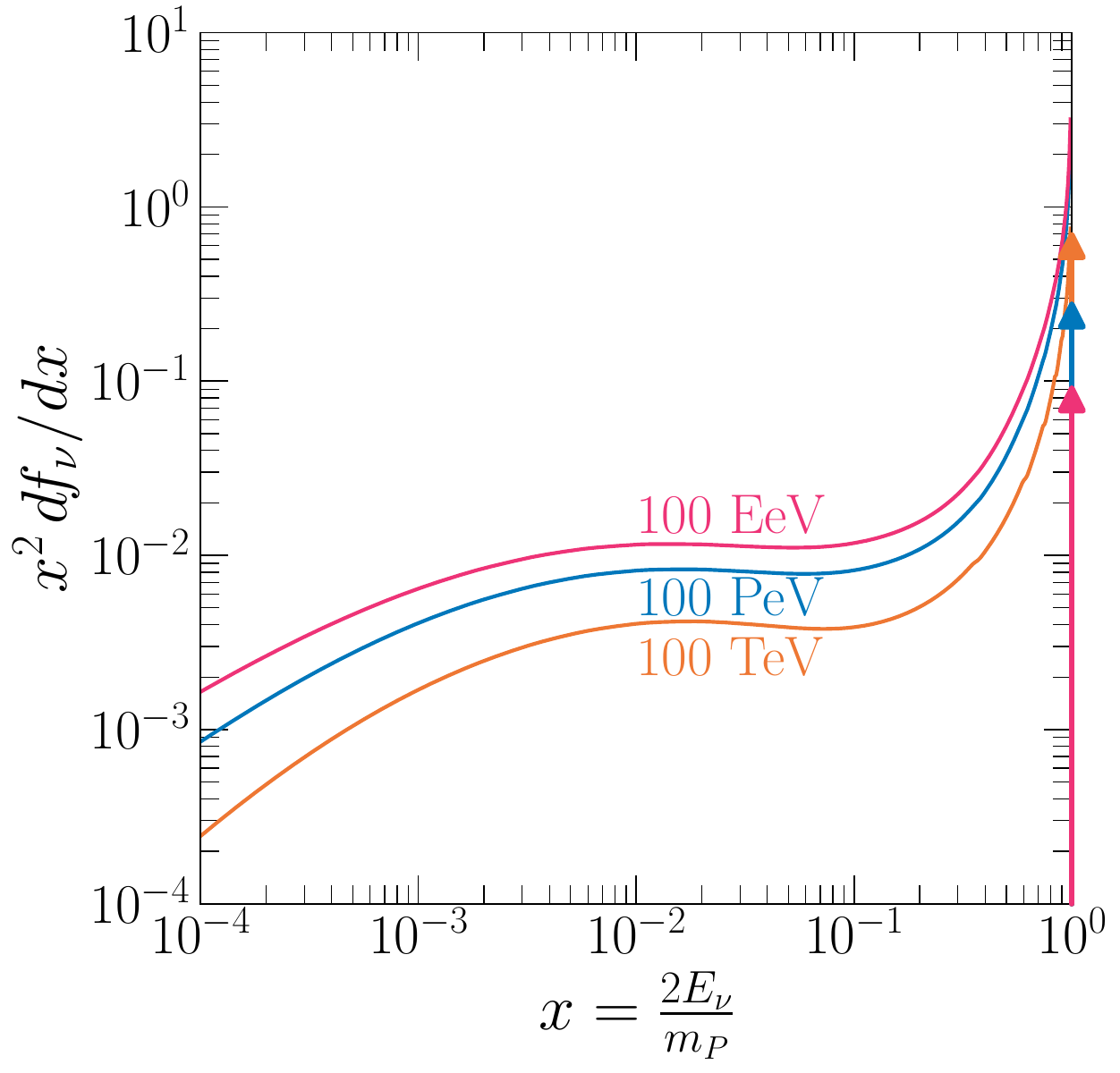}
    \caption{Neutrino spectra at production for a 2-body decay $P\to\nu\nu$ including 
    FSR, computed using \texttt{HDMSpectra}~\cite{Bauer:2020jay}, for parent 
    particle masses of $100$~TeV (orange), $100$~PeV (blue) and $100$~EeV 
    (magenta). The $\delta$-contribution (the fraction of primary neutrinos 
    that do not undergo FSR) is $0.75$, $0.3$, and $0.1$ respectively. The continuous curves show the FSR continuum, while the arrows indicate the $\delta$-contributions.}
    \label{fig:FSR}
\end{figure}

When a high energy neutrino is produced in a two-body decay, it can radiate 
a $W$ or $Z$ boson, transferring part 
of its energy into a gauge boson cascade and producing a continuum of 
secondary neutrinos at lower energies. This final state radiation 
effect grows with energy and cannot be neglected at the masses relevant 
for the KM3NeT event. Figure~\ref{fig:FSR} shows the neutrino spectrum at 
production including FSR, computed using \texttt{HDMSpectra}~\cite{Bauer:2020jay}, 
for three values of the parent particle mass.
The plotted quantity is $x^2\,df_\nu/dx$, where $df_\nu/dx = \Gamma_P^{-1}\, d\Gamma/dx$ is the prompt spectrum per decay, with $x = 2E_\nu/m_P$, $E_\nu$ the neutrino energy, and $\Gamma_P = 1/\tau_P$~\cite{Bauer:2020jay}.
The $\delta$-contribution (the fraction of primary neutrinos that do not undergo FSR) decreases 
from $0.75$ at $100$~TeV to $0.3$ at $100$~PeV and $0.1$ at $100$~EeV, 
confirming that FSR becomes increasingly important at higher energies and 
depletes the monochromatic peak significantly.

\section{Interactions with cosmic neutrinos}
\label{sec:interactions}

\begin{figure}
    \centering
    \includegraphics[width=0.5\linewidth]{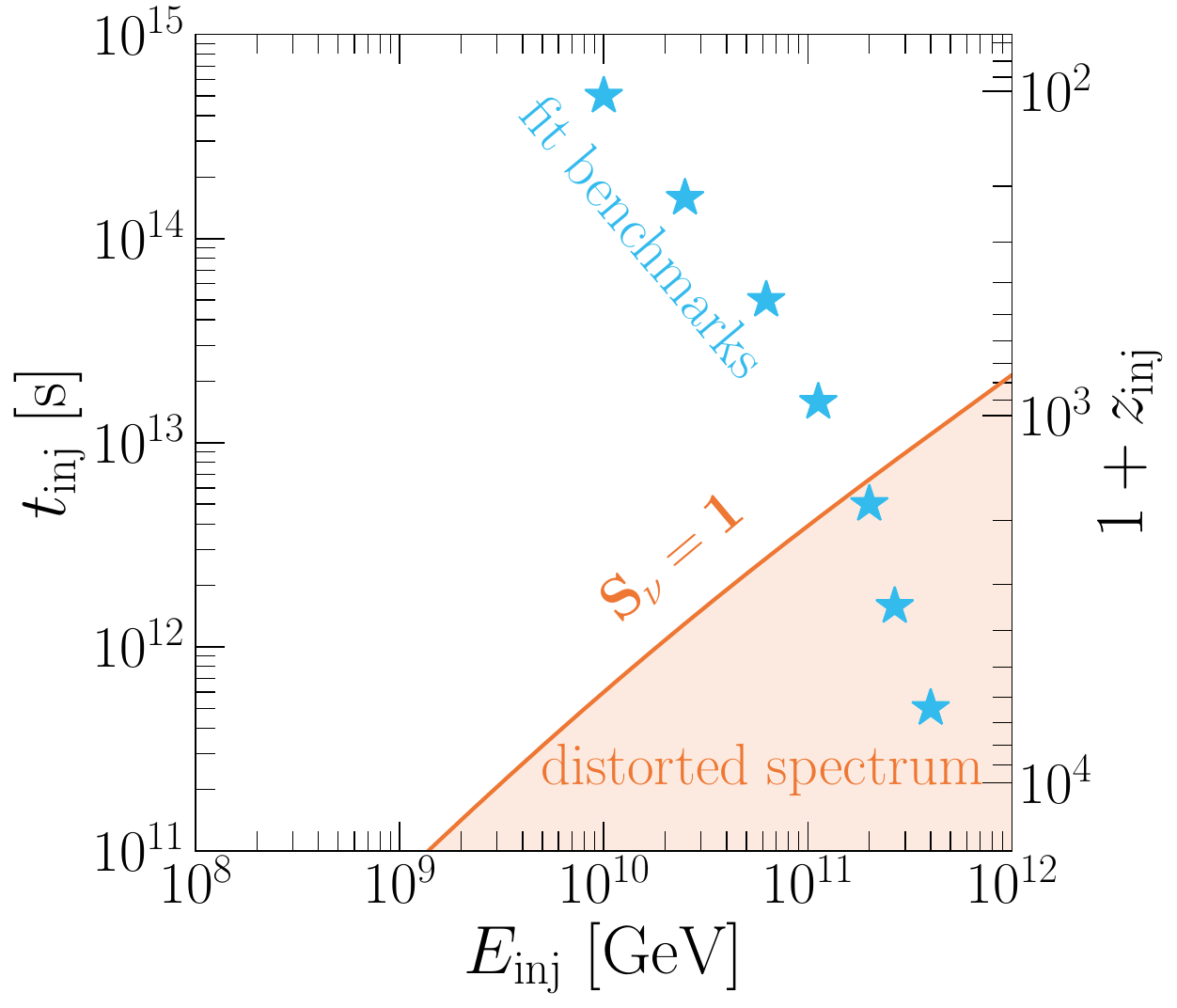}
    \caption{Region where distortion of the Phenu spectrum is expected, as a function of the energy $E_\text{inj}$ and time $t_\text{inj}$ (or redshift $1+z_\text{inj}$, right axis) of injection. The orange line shows $S_\nu = 1$, where $S_\nu$ is the average number of scatterings on the \cnb that a neutrino undergoes between its injection and today (see eq.~(3.2) of ref.~\cite{GrimbaumYamamoto:2025nai}). In the shaded region, $S_\nu > 1$, so that scatterings are expected to distort the spectrum. The stars show the benchmark points of table~\ref{tab:phenu}, at $E_\text{inj} = m_P/2$ and $t_\text{inj} = \tau_P/2$, where the 2-body injection peaks; the benchmarks with $\tau_P = 10^{16}$~s and $10^{17}$~s lie outside the plotted range.}
    \label{fig:scattering}
\end{figure}

In figures~6 and 7 of ref.~\cite{GrimbaumYamamoto:2025nai}, we determined the region of parameter space along which we expect that a Phenu will interact with the \cnb on its way to the detector.
For the KM3NeT event, these figures, a zoomed version of which is shown in figure~\ref{fig:scattering},
show that such scatterings are expected to occur for injection redshifts larger than about 1000, that is to say, for an injection at about the recombination time or before.
To determine the distortion of the neutrino energy spectrum these scatterings induce in this case, it is necessary to develop a dedicated code.

\begin{figure}
  \centering
\begin{tikzpicture}[
  proc/.style = {
    rounded corners=6mm,
    fill=DuskBlue, draw=DuskBlue, line width=0.4pt,
    text=PaperWhite, font=\sffamily\small,
    align=center, text width=68mm, minimum height=11mm, inner sep=3mm
  },
  term/.style = {
    circle,
    fill=DuskBlue, draw=DuskBlue, line width=0.4pt,
    text=PaperWhite, font=\sffamily\small\bfseries,
    align=center, inner sep=2.5mm
  },
  dec/.style = {
    diamond, aspect=2.4,
    fill=LavGrey, draw=DuskBlue, line width=0.4pt,
    text=PaperWhite, font=\sffamily\small\bfseries,
    align=center, inner sep=1mm
  },
  elastic/.style = {
    rounded corners=6mm,
    fill=Fern, draw=DuskBlue, line width=0.4pt,
    text=PaperWhite, font=\sffamily\small,
    align=center, text width=36mm, minimum height=11mm, inner sep=3mm
  },
  inelastic/.style = {
    rounded corners=6mm,
    fill=RosyCopper, draw=DuskBlue, line width=0.4pt,
    text=PaperWhite, font=\sffamily\small,
    align=center, text width=36mm, minimum height=11mm, inner sep=3mm
  },
  arr/.style = {-Stealth, draw=DuskBlue, line width=0.7pt},
  lbl/.style = {font=\sffamily\scriptsize\itshape, text=DuskBlue},
]

\node[proc]  at (0, -2.2)    (inj)   {\textbf{Neutrino injection from relic decays}\\[2pt]{\footnotesize including FSR}};
\node[proc]  at (0, -4.2)    (prop)  {\textbf{Propagation in expanding universe}};
\node[dec]   at (0, -6.4)    (cvb)   {Scattering\\on C$\nu$B?};

\node[elastic]   at (-2.4, -9.2) (elas) {\textbf{Elastic scattering}\\[2pt]{\footnotesize update energy, continue}};
\node[inelastic] at ( 2.4, -9.2) (inel) {\textbf{Inelastic scattering}\\[2pt]{\footnotesize remove neutrino}};

\node[proc]  at (0, -11.8)   (red)   {\textbf{Redshift energy loss}\\[2pt]{\footnotesize and flavour oscillations}};
\node[dec]   at (0, -14.0)   (z0)    {$z=0$?};
\node[term]  at (4.5, -14.0) (end)   {Spectrum\\at Earth};

\draw[arr] (inj)   -- (prop);
\draw[arr] (prop)  -- (cvb);
\draw[arr] (red)   -- (z0);

\draw[arr] (z0.east) -- (end.west)
  node[lbl, above, midway] {Yes};

\coordinate (yespt) at (0, -7.8);
\draw[DuskBlue, line width=0.7pt] (cvb.south) -- (yespt)
  node[lbl, right, pos=0.5] {Yes};
\draw[arr] (yespt) -| (elas.north);
\draw[arr] (yespt) -| (inel.north);

\draw[arr] (elas.south) |- (red.west);
\draw[arr] (inel.south) |- (red.east);

\draw[arr] (cvb.east) -- ++(4.5,0) |- (red.east);
\node[lbl, above right, xshift=2mm] at (cvb.east) {No};

\draw[arr] (z0.west) -- ++(-4.5,0) |- (inj.west);
\node[lbl, above left, xshift=-10mm] at (z0.west) {No};

\end{tikzpicture}
  \caption{Flowchart of the Monte Carlo evolution of the Phenu spectrum }
  \label{fig:flowchart}
\end{figure}
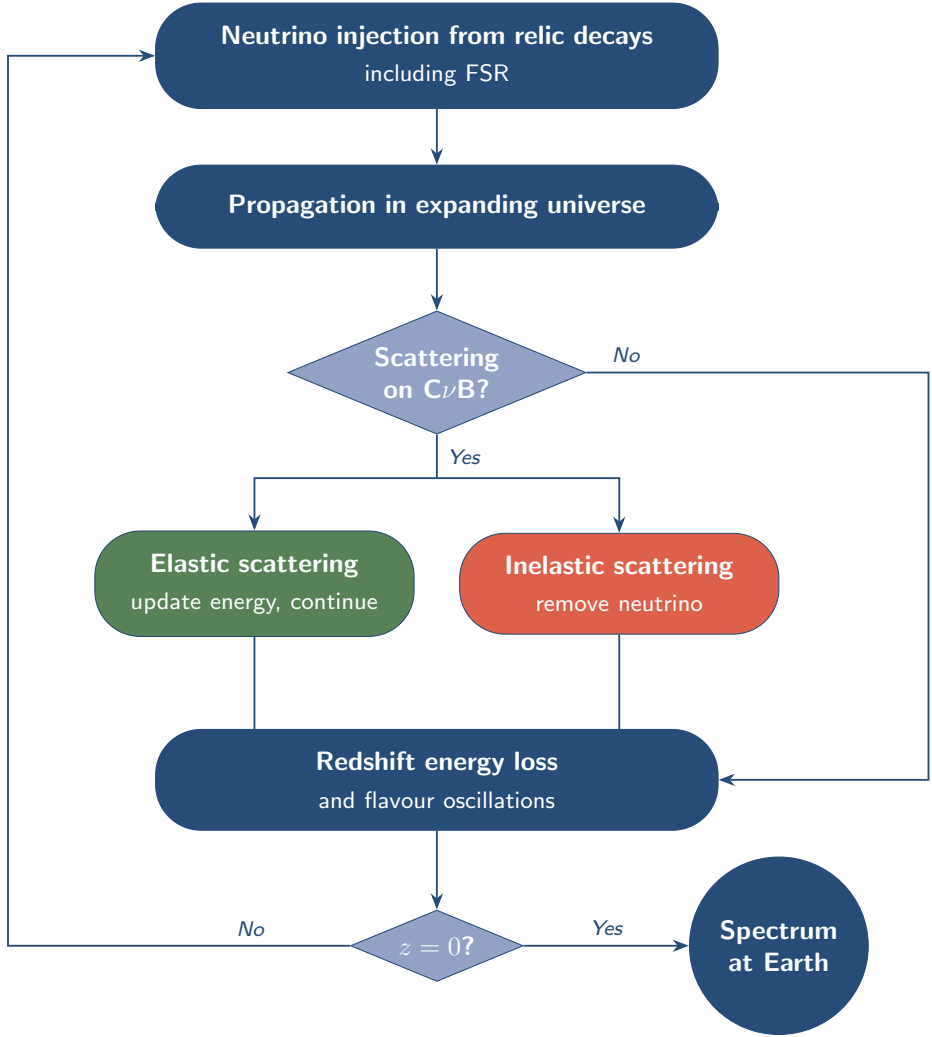

The structure of the code we developed is given in the chart of figure~\ref{fig:flowchart}.
This code is not based on an integration of the coupled Boltzmann equations determining the 
temporal evolution of the Phenu flux and spectrum, which is difficult to do in an efficient way.
Instead, in a spirit similar to other codes that have been recently developed for other purposes~\cite{Bianco:2025boy, DiMarco:2025qby, Ovchynnikov:2024rfu, Ovchynnikov:2024xyd}, we adopted 
a Monte Carlo inspired 
probabilistic approach that tracks the individual neutrino energies, all the way from the time of 
their emission to today.
Such an approach allows for an exact treatment of the kinematics 
at each interaction. The 
simulations used in the present work follow the decay of $10^5$ parent particles ($10^4$ in the case of $\tau_P = 10^{12}$s), producing a 
final sample of approximately $10^{\text{6-8}}$ neutrinos across all six 
species in the detector ($\nu_e, \nu_\mu, \nu_\tau$ and their antineutrinos), which are 
tracked separately throughout the evolution.

The time evolution begins at the redshift $z(t = \tau_P/100)$, 
corresponding to a cosmic time well before the characteristic decay 
epoch, so that essentially no decays have occurred before. 
It is then discretised over small adaptive redshift steps $dz$, defined 
such that the characteristic times between
two Phenu interactions 
 are large in comparison. At every step, we compute the probability for 
the relics to decay into neutrinos,
\begin{equation}
    dP_\text{decay} = 1 - e^{-dt/\tau_P},
\end{equation}
taking into account FSR through the spectra provided by 
\texttt{HDMSpectra}~\cite{Bauer:2020jay}. 
At each step, one computes the probability that each neutrino in the sample
undergoes either elastic, inelastic, or no 
scattering.\footnote{We refer to `elastic' 
interactions as those that do not change the number of neutrinos, even 
if they change the flavour, such as $\nu_i\bar{\nu}_i^\text{BG} \to 
\nu_j \bar{\nu}_j$ or $\nu_i\nu_i^\text{BG} \to \nu_j\nu_j$.} Each interaction is sampled from the relevant cross 
sections: elastic cross sections are taken at tree level from ref.~\cite{GrimbaumYamamoto:2025nai}, while inelastic ones are tabulated at leading order with \texttt{MadGraph}~\cite{Alwall:2014hca} and interpolated.
Both depend on the Mandelstam variable 
$s = 2EE_\text{BG}(1-\cos\theta)$, where $E_\text{BG}$ is the energy 
of the background neutrino, sampled from a Fermi-Dirac distribution at 
temperature $T_\text{BG}(z) = (4/11)^{1/3}T^0_\text{CMB}(1 + z)$, and 
$\theta$ is the angle between the two neutrinos.\footnote{We neglect the neutrino masses in the \cnb distribution. The background neutrinos become non-relativistic only at $1+z \lesssim 200$, whereas the scattering rate relative to the Hubble rate decreases rapidly with time, so that virtually all scatterings occur well before this for the lifetimes considered.}
Although the \cnb\ is isotropic, the probability that a given background neutrino scatters off the Phenu is not uniform in $\theta$, since $s$ itself depends on it.
We therefore sample $\theta$ from the differential interaction rate.

In the case of an elastic scattering, the background neutrino is 
upscattered by the Phenu, leading to two final-state high energy 
neutrinos, which must be and are tracked afterwards. The outgoing neutrino energies are obtained by 
boosting the scattering into the centre-of-mass frame, sampling the outgoing angle from 
the differential cross sections of ref.~\cite{GrimbaumYamamoto:2025nai}, and boosting back into 
the cosmic frame. In contrast, for inelastic scatterings, both the 
background and high energy neutrinos are lost to other Standard Model 
particles that eventually cascade down to secondary neutrinos of much 
lower energy. For the present 
work, those secondary neutrinos are irrelevant as their energies are far below 
the high energy peak of the distribution. Thus we make the ansatz that 
whenever an inelastic interaction occurs, we remove the Phenu that disappears in the scattering as well as the secondary neutrinos this Phenu produces.

After determining whether a scattering has occurred at a given redshift and 
accounting for its effect if it did, we redshift the whole energy spectrum by 
a factor $(1+z_\text{new})/(1+z_\text{old})$ (where $z_\text{new} = z_\text{old} - dz$) before proceeding to the next step.
Neutrinos produced in a definite flavour, i.e.~the primary neutrinos and those from the FSR decays of charged leptons and mesons, lead after oscillations to an averaged flux whose composition is given by the oscillation matrix
\begin{align}\label{eq:osc}
    (\mathbf{M}_{\text{osc}})_{\alpha\beta}=\sum_{i=1}^3 
    |U_{\alpha i}|^2|U_{\beta i}|^2\approx
    \begin{pmatrix} 
    0.55&0.17&0.28\\
    0.17&0.45&0.37\\
    0.28&0.37&0.35
    \end{pmatrix}\;,
\end{align}
applied independently to neutrinos and antineutrinos, with $U$ the PMNS matrix.
This is justified as long as the oscillations average out before the next scattering, i.e.~as long as the oscillation time, $t_\text{osc} = 4\pi E/\Delta m^2_{21} \simeq 0.11~\text{s}\,(E/\text{GeV})$~\cite{Esteban:2024eli}, is shorter than the time between two scatterings, which holds for $\tau_P \gtrsim 3\times10^{12}$~s.
For shorter lifetimes, the neutrinos are injected at higher redshift, where the interaction rate is larger at fixed $m_P$.
Indeed, the \cnb number density grows as $(1+z)^3$, while $s$, and hence the cross section below the $Z$ resonance, grows as the \cnb temperature, i.e.~as $(1+z)$, so that neutrinos can scatter before the oscillations average out.
These scatterings, however, are flavour blind to a very good approximation, so that they do not affect the flavour composition.
Elastic scatterings proceed through $Z$ exchange, which is flavour universal and leaves the mass state composition of the scattered neutrino unchanged, so that no oscillation is applied after them.
Similarly, the total inelastic rate is the same for the three flavours to better than $0.2\%$ wherever scatterings are frequent.
Larger differences, of up to about $35\%$, only arise at lower centre-of-mass energies, but this is relevant only where the interaction rate is low, so that this affects the flavour composition at the percent level at most.
Note that the channel $\nu_l\bar\nu_l \to \nu_{l'}\bar\nu_{l'}$ and the upscattered background neutrinos produce all flavours equally, which drives the composition of the scattering products towards the democratic one.

The process is then repeated until today, giving the per-flavour energy spectrum to be observed in the detector. For definiteness, in this work we consider the decay $P\to \nu_e \bar\nu_e$ but due to FSR and oscillations, the electron flavour is redistributed across all three flavours. 
The flavour composition of the bulk of the spectrum is set at production by the FSR shower, whose neutrinos come mostly from pion and muon decays, and is close to democratic, $\nu_e:\nu_\mu:\nu_\tau \simeq 0.30:0.36:0.34$, for all masses and lifetimes considered.
The peak, dominated by primary $\nu_e$ that radiated little, has instead $\nu_e:\nu_\mu:\nu_\tau \simeq 0.55:0.17:0.28$, i.e.~the $\nu_e$ row of $\mathbf{M}_{\text{osc}}$.
For the shortest lifetimes, elastic scatterings slightly pull the transition region between the bulk and the peak of the Phenu spectrum towards the democratic composition.
The impact of this non-democratic composition of the peak on the fit is discussed in section~\ref{sec:fit}.

\section{Characterisation of the Phenu flux}
\label{sec:characterisation}

\begin{figure}
    \centering
    \includegraphics[width=\textwidth]{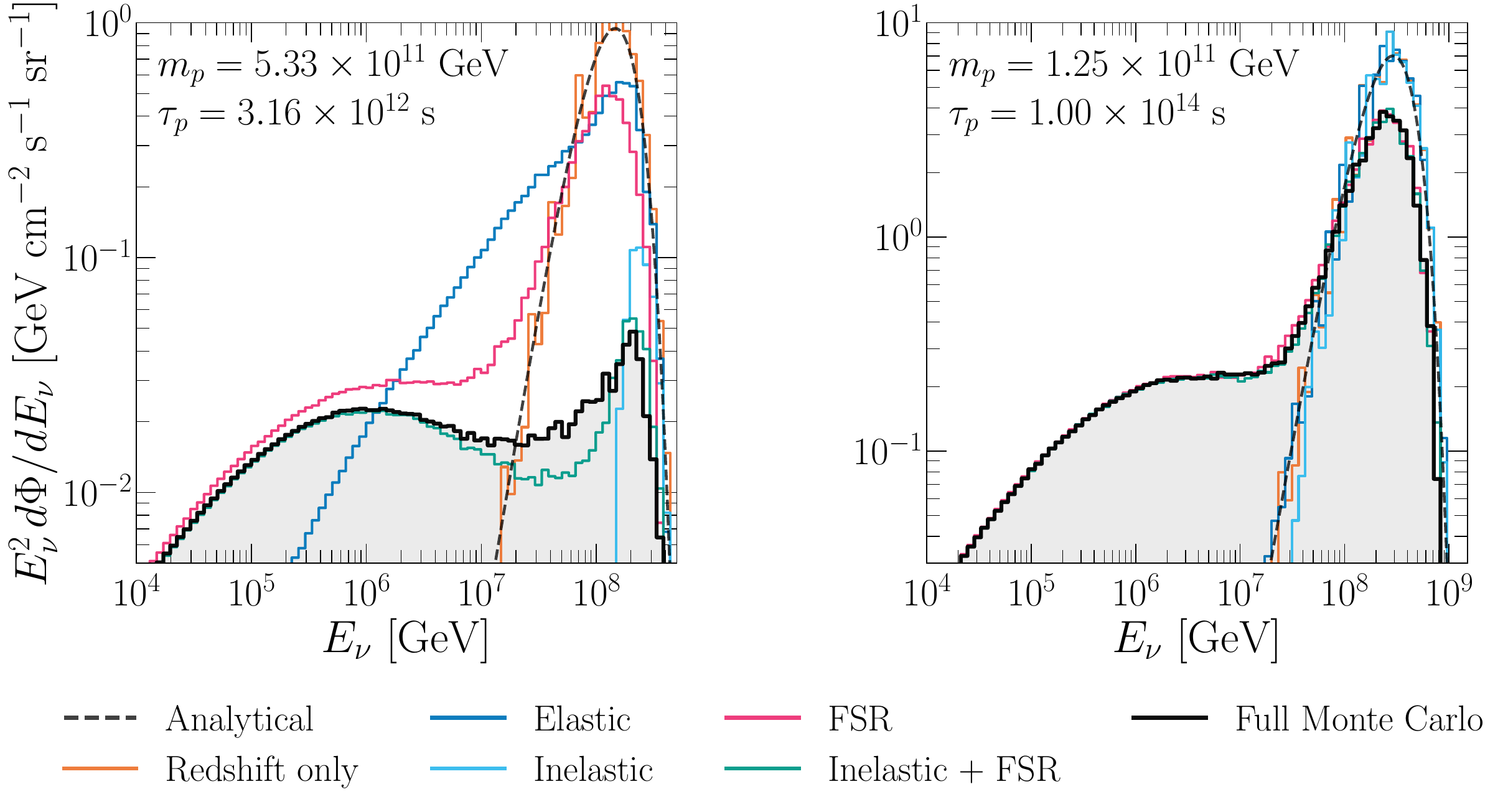}
    \caption{Comparison of the various effects impacting the Phenu spectrum (summed over all flavour neutrinos and antineutrinos), compared to the analytical two-body decay spectrum from~\cite{GrimbaumYamamoto:2025nai} for $m_P = 5.33\times 10^{11}$ GeV, $\tau_P = 3.16 \times 10^{12}$ s, $f_P = 1$ (left) and for $m_P = 1.25 \times 10^{11}$ GeV, $\tau_P = 10^{14}$ s, $f_P = 1$ (right).}
    \label{fig:comparison}
\end{figure}

In the absence of FSR effects, the primary neutrinos produced 
by the decay of $P$ carry an energy equal to half the mass of the source particle, giving rise to a sharp 
monochromatic spectral feature at the source. 
Since all source particles do not decay at the same time, so that not all Phenus are redshifted in the same way, the spectrum in the detector is not monochromatic anymore. Without interactions with the \cnb, the spectrum follows from eq.~(2.13) of ref.~\cite{GrimbaumYamamoto:2025nai} with the prompt spectrum $df_\nu/dE_\text{inj} = \delta(E_\text{inj} - m_P/2)$ and $\Delta n_\nu = 2$:
\begin{equation}\label{eq:dNdE_2body}
    \frac{d N_\nu}{dE_\nu} = \frac{2\,\Omega_P^0\,\rho_{\text{crit}}^0}
    {m_P H_\text{inj} \tau_P E_\nu} 
    e^{-t_{\text{inj}}/\tau_P}\,
    \Theta\!\left(\frac{m_P}{2} - E_\nu\right),
\end{equation}
which is still quite sharp,
with $t_{\text{inj}}$ corresponding to the injection time of the neutrinos that are observed with a given energy $E_\nu$ today.\footnote{The injection time is equal to $1/(2H_{\text{inj}})$ and $2/(3H_{\text{inj}})$ for an injection during the radiation or matter dominated era, respectively, with the Hubble constant at injection given by $H_{\text{inj}}=m_P^2/(8t_rE_\nu^2)$ and $H_{\text{inj}}=\frac{2}{3t_m}(m_P/2E_\nu)^{3/2}$, and $t_r=2.4\cdot 10^{19}$~s, $t_m=5.5\cdot 10^{17}$~s, respectively, see~\cite{GrimbaumYamamoto:2025nai}. 
The peak energy of this feature is given by
    $E_{\text{peak}} \simeq \frac{m_P}{2} \sqrt{\frac{\tau_P}{2t_r}}$ and $E_{\text{peak}} \simeq \frac{m_P}{2}(\tau_P/3t_m)^{2/3}$ respectively.}
Here $dN_\nu/dE_\nu = 4\pi\,d\Phi/dE_\nu$ is the flux per unit energy integrated over solid angle, with $d\Phi/dE_\nu$ the flux per unit energy and solid angle, $\Omega_P^0$ is the fraction of the critical energy density today, $\rho_\text{crit}^0$, that the source particles would have if they had not decayed, and we define the abundance $f_P \equiv \Omega_P^0/\Omega_\text{DM}^0$.
As a validation of the code, we first verified that in the absence of FSR and \cnb interactions, the simulated spectrum reproduces well this analytical spectrum. This spectrum is shown by the orange line in figure~\ref{fig:comparison} for two lifetime values, taking in each case $m_P$ such that $E_{\text{peak}}$ matches the reconstructed 
energy of the KM3-230213A event (see table~\ref{tab:phenu} below).

On top of showing the spectra in the detector for the purely redshifted case, figure~\ref{fig:comparison} also shows the spectra we get by adding various effects: FSR only, elastic scatterings only, inelastic scatterings only, inelastic with FSR, and the full Monte Carlo spectrum.
For the first lifetime, $\tau_P=3.2\cdot 10^{12}$~s, the interactions with the \cnb turn out to be important, unlike for the other case, $\tau_P= 10^{14}$~s, where these effects are moderate.
The FSR curve directly follows from the spectrum at source with FSR, figure~\ref{fig:FSR}, when the redshift effects are added. 
Thus, as expected from this figure, FSR depletes the high energy peak, corresponding to the energy lost of the primary neutrinos radiating electromagnetic (EM) and hadronic showers that produce large numbers of lower energy secondary neutrinos. 
Elastic scatterings also deplete the high energy peak and populate the relatively lower energy neutrino spectrum, since each interacting Phenu upscatters a \cnb neutrino into two lower energy neutrinos with about half the incoming Phenu energy. 
Finally, inelastic scatterings deplete the overall neutrino spectrum, with a stronger effect on neutrinos emitted earlier, which correspond to lower energies today, resulting in a sharpening of the spectral shape of the signal. 
This suppression effect of the inelastic scatterings, resulting in a sharpening of the spectrum, is in agreement with the general expectations of~\cite{Gondolo:1991rn}. 
This sharpening effect is nevertheless attenuated by the elastic scatterings that induce an opposite effect.
The spectrum inelastic + FSR, i.e.~without elastic scatterings, corresponds to the spectrum of neutrinos that did not interact during the evolution.
This follows from our ansatz that neutrinos that interact inelastically are removed from the simulation (together with the final state products of these interactions), and elastic scatterings being turned off, meaning the resulting spectrum only shows the depletion of the peak due to FSR and inelastic scatterings.
Thus, the comparison of this line with the full result line shows that, as expected, the elastic scatterings repopulate the spectrum at energies a few times lower than at the peak.
Note also that the total spectrum we find is in qualitative agreement with the spectrum that had been obtained~\cite{Ema:2014ufa} from a Boltzmann equation approach for PeV energy neutrinos (in the context of the high energy neutrinos observed by IceCube).

All in all, the final spectrum obtained is sharper than the original one but the height of the peak can be sizeably suppressed as soon as the lifetime is shorter than about the recombination time ($t_{rec}=10^{13}$~s). For instance, for $\tau_P=3.2\times10^{12}$~s (left panel of figure~\ref{fig:comparison}) the suppression factor is $\sim10$. For even shorter lifetimes, the high energy peak quickly becomes completely erased by the scatterings on \cnb neutrinos.
In this latter case, the spectrum is instead dominated by a 
broad, elevated bump of secondary neutrinos at lower energies, mostly coming from FSR. 
For post-recombination lifetimes, for instance for $\tau_P= 10^{14}$~s (right panel), the peak is still quite prominent, as its height is suppressed only by a factor $\sim 2$, and the amount of lower energy secondary neutrinos is much smaller.\footnote{The fact that this transition between both regimes occurs for lifetimes close to the recombination time appears to be a pure coincidence, as both times rely on different physical principles.}
Thus, a precise measurement of the shape of the spectrum at the peak would allow one to determine the lifetime, mass and abundance of the source particle, see figure~\ref{fig:comparison}.

The effect of FSR at each scattering on the \cnb, which is not included in our code, is discussed in appendix~\ref{app:fsr_scattering}.

\section{Fit to the data}
\label{sec:fit}


\subsection{Datasets}
\paragraph{UHE datasets}

The fits we will perform cover different energy ranges corresponding to different experimental data samples of KM3NeT, IceCube and Auger. For ultrahigh energy (UHE) events these ranges are
\begin{equation}\label{eq:intervals}
\left\{
\begin{aligned}
\mathcal{I}^\text{KM3}_\text{in} &= [72\,\mathrm{PeV}, 2.6\,\mathrm{EeV}]\,, \\
\mathcal{I}^\text{KM3}_\text{out} &= [100\,\mathrm{TeV}, 100\,\mathrm{EeV}]\setminus \mathcal{I}^\text{KM3}_\text{in}\,,  \\
\mathcal{I}^\text{Auger} &= [10\,\mathrm{PeV}, 1\,\mathrm{ZeV}]\,, \\
\mathcal{I}^\text{IC} &= [20\,\mathrm{PeV}, 50\,\mathrm{EeV}]\,.
\end{aligned}
\right.
\end{equation}
Within these ranges, only one event has been recorded, the KM3-230213A event that has been observed within the interval $\mathcal{I}^\text{KM3}_\text{in}$.\footnote{This interval corresponds to the central 90\% energy range of simulated neutrino events producing the observed muon in ARCA, assuming a single power law with spectral index $\gamma=2$ \cite{KM3NeT:2025npi}. Strictly speaking, this range should be recomputed for each fitted spectral model, which would require full access to the detector simulation.} 
The samples used (effective areas, livetime, ...) are detailed in appendix~\ref{app:statistics}. A democratic 1:1:1 flavour composition is considered at the detector.\footnote{Since only the flavour-averaged effective area of KM3NeT is available, we assume a democratic flavour composition for the Phenu flux.
Our Monte-Carlo code shows that it deviates from this at the peak of the spectrum, where the $\nu_\mu$ fraction is only about half the democratic one (see section~\ref{sec:interactions}).
We redid the analysis assuming that KM3NeT is sensitive to $\nu_\mu$ only, or to $\nu_\mu$ and $\nu_\tau$ only, as the event selection is based on tracks.
We also reweighted the Auger exposure channel by channel and kept IceCube-EHE flavour-democratic.
Both the irreducible tension and the tension of the best Phenu fits increase, by about $0.1$--$0.3\sigma$.
The best fits lie closer to the irreducible tension, within $0.03$--$0.07\sigma$ instead of $0.1\sigma$, so that our conclusions are unchanged.}

\paragraph{IceCube HESE sample}

Below we will also perform a fit that includes the (diffuse astrophysical) neutrino events that have been observed at lower energies by the IceCube observatory, in the HESE 7.5-year data sample~\cite{IceCube:2020wum}, see more details in appendix~\ref{app:statistics} too.

\subsection{Compatibility of UHE datasets}

We follow the Bayesian statistical framework of ref.~\cite{KM3NeT:2025ccp} (see more details in appendix~\ref{app:statistics}) to assess the compatibility of the KM3-230213A observation with the non-observations by IceCube-EHE, the extremely high energy (EHE) selection, and Pierre Auger at the KM3NeT event energy, reproducing their results as a validation of our implementation.
In the following, the best-fit values of the parameters are indicated by a hat.

\subsubsection{\texorpdfstring{Joint fit of a power law within the UHE KM3NeT interval $\mathcal{I}^\text{KM3}_\text{in}$}{Joint fit of a power law within the UHE KM3NeT "in" interval}}\label{sec:fit_phenu_E2}

As a first check, we assume a fixed $E^{-2}$ spectrum, i.e.~we assume a per-flavour and $\nu$-$\bar{\nu}$ averaged flux of the form $E^2\Phi^\text{1f}_{\nu+\bar\nu}(E)\equiv \phi_0$, and determine the best-fit value for $\phi_0$ by maximising the likelihood for KM3NeT, IceCube and Auger datasets over the energy interval $\mathcal{I}^\text{KM3}_\text{in}$.
The best-fit is $\hat\phi_0 = 5.55^{+7.54}_{-3.88} \times 10^{-10}~\mathrm{GeV\,cm^{-2}s^{-1}sr^{-1}}$, and the probability of observing one event in KM3NeT and zero in both IceCube-EHE and Auger at this best-fit is $3.5\times10^{-3}$ or, equivalently, a tension of $2.7\sigma$ between the datasets.
This is to be compared with $7.5^{+13.1}_{-4.7}\times10^{-10}~\mathrm{GeV\,cm^{-2}s^{-1}sr^{-1}}$ and a tension of $2.6\sigma$ in ref.~\cite{KM3NeT:2025ccp}.
The difference in normalisation follows from our use of the updated 12.6-year IceCube-EHE exposure rather than the 9-year analysis they employ. \footnote{Adopting their 9-year exposure we recover $\hat\phi_0 = 7.12\times10^{-10}~\mathrm{GeV\,cm^{-2}s^{-1}sr^{-1}}$, in agreement with their value at the 5\% level.}

We then reproduce the single power law (SPL) fit assuming a parameterisation
\begin{equation}\label{eq:SPL}
    \Phi^\text{1f}_{\nu+\bar\nu}(E) \equiv \phi_0 \left(\frac{E}{100 \text{ TeV}}\right)^{-\gamma}\,,
\end{equation}
where $\phi_0$ and $\gamma$ represent the flux normalisation and the spectral index.
We compute the likelihood $\mathcal{L}^\text{in}_\text{UHE} =\prod_d \mathcal{L}^\text{in}_d$ within the energy interval $\mathcal{I}_\text{in}^\text{KM3}$. 
To assess the compatibility of the UHE datasets, the Bayesian approach we perform uses the posterior predictive check (PPC) method~\cite{gelman1996posterior}, with the posterior $P(\Theta=\{\phi_0, \gamma\}) \equiv \mathcal{L}^\text{in}_\text{UHE}(\phi_0, \gamma) \times \mathcal{U}_{[0, 10^{-10}]}(\phi_0) \times\mathcal{U}_{[1, 4]}(\gamma)$ where $\mathcal{U}_{[a,b]}(\theta)$ is a continuous uniform prior between $a$ and $b$ on the parameter $\theta$. The joint probability is~\cite{KM3NeT:2025ccp}
\begin{equation}\label{eq:ppc}
    p_\text{PPC} = \int\sum_{N_\text{rep}|p(N_\text{rep}|\Theta)\leq p(N_\text{obs}|\Theta)} p(N_\text{rep}| \Theta)P(\Theta)d\Theta,
\end{equation}
where $N_\text{obs}$ and $N_\text{rep}$ are, respectively, the observed number of events in each sample and the number of events in replicated toy data following the posterior $P(\Theta)$, and $p(N_\text{rep}|\Theta)$ is the Poisson probability of this replicated data assuming parameters $\Theta$.
We obtain a final $z$ score $z_\mathrm{PPC} = 2.01\sigma$ using the one-tailed convention, to be compared with the $1.8\sigma$ of ref.~\cite{KM3NeT:2025ccp} for the same fit. \footnote{As an independent check we also perform the frequentist parameter-goodness-of-fit as in ref.~\cite{KM3NeT:2025ccp}, obtaining $z_\mathrm{PG} = 2.6\sigma$ against their $2.5\sigma$.}
Both residual discrepancies are attributable to the updated IceCube-EHE effective area and 12.6-year exposure time.

Due to the narrow energy range of $\mathcal{I}^\mathrm{KM3}_\mathrm{in}$, 
the UHE-only SPL likelihood is degenerate in the $(\phi_0, \gamma)$ 
plane and does not yield meaningful individual constraints on the two 
parameters. Indeed, any SPL parameterisation that produces the same integrated flux within that bin will fit equally well. Best-fit values are therefore only quoted after incorporating 
the HESE likelihood as performed in the next section, which provides the necessary lever arm in energy 
to break this degeneracy.

\subsubsection{Joint global fit of a power law within the UHE and HESE energy ranges}

Next we extend the analysis by performing a fit over all four UHE ranges of Eq.~\eqref{eq:intervals}, also
incorporating the IceCube HESE 7.5-year
measurement~\cite{IceCube:2020wum}. Here the likelihood is computed as
\begin{equation}\label{eq:global_likelihood}
    \mathcal{L}^\text{(HESE)}_\mathrm{SPL}(\phi_0,\gamma) = \mathcal{L}_\text{UHE}(\phi_0,\gamma)\times\mathcal{L}^\text{(HESE)}_\text{IC}(\phi_0, \gamma)\,,
\end{equation}
where $\mathcal{L}_\text{IC}^\text{(HESE)}$ is extracted from the IceCube public data release associated with  \cite{IceCube:2020wum}.
The Bayesian analysis is performed using a posterior $P^\text{(HESE)}(\phi_0, \gamma)$ defined as the product of the likelihood from eq.~\eqref{eq:global_likelihood} and the uniform priors $\mathcal{U}_{[0, 4\times 10^{-18}]} (\phi_0)$ (in $\mathrm{GeV^{-1}s^{-1}cm^{-2}sr^{-1}}$) and $\mathcal{U}_{[1.5, 3.5]}(\gamma)$, the latter chosen to match the range over which the public HESE scan is defined.\footnote{Ref.~\cite{KM3NeT:2025ccp} quotes a prior $\mathcal{U}_{[1,4]}(\gamma)$, while the scan it relies on is tabulated only for $\gamma\in[1.5,3.5]$.}
The best-fit flux is obtained by finding the maximum $[\hat\phi_0, \hat\gamma = \arg\max_{(\phi_0, \gamma)}P^\text{(HESE)}(\phi_0, \gamma)]$, and the highest-posterior-density (HPD) 1$\sigma$ intervals are extracted from marginalised posteriors,
\begin{equation}
    \hat\phi_0 = 1.90^{+0.50}_{-0.51}
    \times 10^{-18}~\mathrm{GeV^{-1}cm^{-2}s^{-1}sr^{-1}},
    \qquad \hat\gamma = 2.73^{+0.17}_{-0.15},
\end{equation}
in good agreement with the results of ref.~\cite{KM3NeT:2025ccp}.
The PPC is done using eq.~\eqref{eq:ppc} with the posterior $P^\text{(HESE)}$ to get the probability of the UHE observations and yields $z_\mathrm{PPC} = 2.56\sigma$, indicating, as expected, that the inclusion of the HESE datasets aggravates the tension with KM3NeT for a power law.

\subsection{Joint fit with a Phenu component}\label{sec:fit_phenu}

In addition to considering a power law, necessary to account for the observation of neutrino events at lower energies ($\sim$~PeV), we also consider a Phenu component, so that the total flux  is
\begin{equation}
\label{eq:flux_fit}
    \Phi^{1f}_{\nu+\bar\nu}(E;\phi_0,\gamma,f_P, \tau_P, m_P)
    = \phi_0\!\left(\frac{E}{100~\mathrm{TeV}}\right)^{-\gamma}
    + f_P \times \Phi^\mathrm{Phenu}_\mathrm{norm}(E;\tau_P,m_P),
\end{equation}
where $\Phi^\mathrm{Phenu}_\mathrm{norm}$ is the Phenu spectrum from
our Monte Carlo simulation for a relic of mass $m_P$ and lifetime
$\tau_P$, normalised to unit abundance, and $f_P \equiv \Omega_P^0/\Omega^0_\text{DM}$ is the relic abundance normalised to that of dark matter (evaluated today, assuming the source particles had not decayed).
Fixing the parameters so that the high energy peak lies at the KM3NeT event energies, there are now 2 additional free parameters beyond $(\phi_0,\gamma)$, that can be taken to be $\tau_P$ and $f_P$.
We select 9 benchmark values for $\tau_P$ ranging from $10^{12}$s to $10^{17}$s and run the likelihood analysis sequentially for each lifetime, so that the posterior is a function of 3 parameters only.

Since the Phenu spectrum peaks at $E_\nu^\mathrm{max}\sim10^8$~GeV, well
above the HESE sensitivity range, the Phenu component contributes
negligibly to $\mathcal{L}^\mathrm{(HESE)}_\mathrm{IC}$ for most
spectra, and in this case the log-likelihood factorises as
\begin{equation}
    \ln\mathcal{L}(\phi_0,\gamma,f_P)
    = \ln\mathcal{L}_\mathrm{UHE}(\phi_0,\gamma,f_P)
    + \ln\mathcal{L}^\mathrm{(HESE)}_\mathrm{IC}(\phi_0,\gamma),
\end{equation}
with expected counts
$\mu^d_\text{exp}(\mathcal{I};\phi_0, \gamma, f_P, \tau_P) = \mu^d_\mathrm{SPL}(\mathcal{I};\phi_0,\gamma) +
f_P\,\mu^d_\mathrm{Phenu}(\mathcal{I};\tau_P)$.
In practice the HESE scan strongly dominates the inference on
$(\phi_0,\gamma)$, so the factorisation is an excellent approximation
for most spectra (see Table~\ref{tab:phenu} below where the
best-fit $(\phi_0,\gamma)$ values coincide with the HESE-only
best-fit).\footnote{The best-fit values from the 7.5 year HESE IceCube analysis are given for an all-flavour flux, so $\hat{\phi}_0$ should be multiplied by a factor of 3 to recover their results.} For the shortest-lifetime spectrum ($\tau_P = 10^{12}$~s), the Phenu flux has a non-negligible
contribution at PeV energies, breaking the factorisation and
shifting $\hat\gamma$. This also happens to a lower extent for the second shortest-lifetime one ($\tau_P =3.16\times10^{12}$~s).

A three-dimensional Bayesian posterior over $(\phi_0, \gamma,
\log_{10}f_P)$ is evaluated with flat priors, with the $\log_{10}f_P$
grid bounded by $f_P\,\mu^\mathrm{KM3}_\mathrm{Phenu,in} \in
[10^{-4}, 5]$ (i.e. such that the expected number of events due to a Phenu at KM3NeT, within the energy interval $\mathcal{I}_\text{in}$, spans from $10^{-4}$ to 5), and the priors on $\phi_0$ and $\gamma$ remain as in the previous fit.

\begin{figure}
    \centering
    \includegraphics[width=0.95\linewidth]{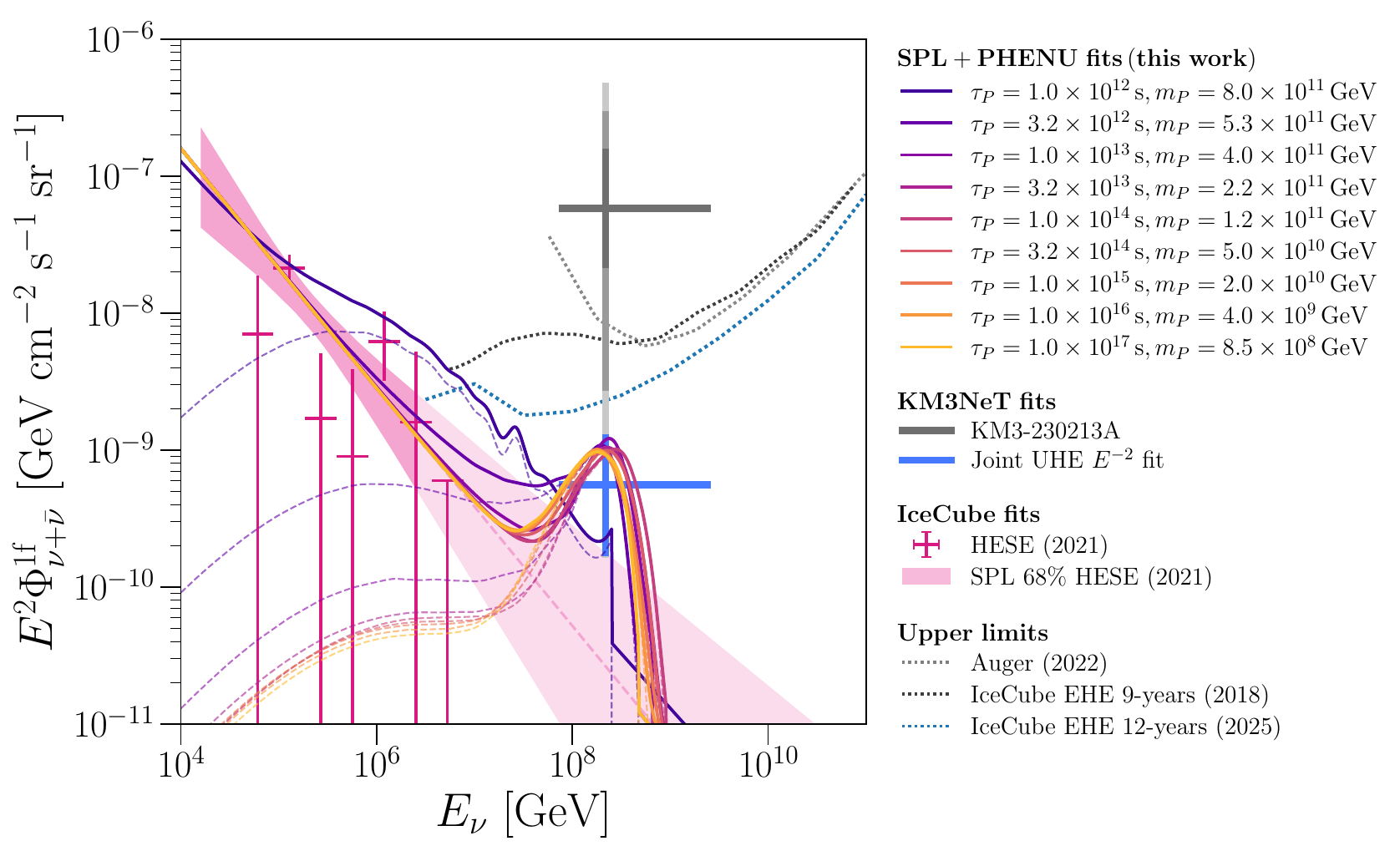}
    \caption{Best-fit Phenu + SPL flux for various lifetimes $\tau_P$ to the KM3-230213A event~\cite{KM3NeT:2025npi}, the Pierre Auger~\cite{PierreAuger:2019azx} and IceCube EHE~\cite{IceCubeCollaborationSS:2025jbi} limits, and the IceCube HESE data~\cite{IceCube:2020wum}. The corresponding parameter values are given in Table~\ref{tab:phenu}. The dashed curves show the Phenu component of the flux. The pink dashed line shows the HESE SPL best-fit. Also shown in blue is the joint ultrahigh energy (Auger, EHE and KM3NeT) best-fit value of section~\ref{sec:fit_phenu_E2} assuming an $E^{-2}$ single power law. }
    \label{fig:results}
\end{figure}

The results are summarised in Table~\ref{tab:phenu} and the corresponding spectra are shown in figure~\ref{fig:results}, alongside the UHE limits from Auger, IceCube as well as the HESE datapoints and SPL best-fit.
All nine spectra reduce the tension observed in the UHE datasets, with eight of them giving $z_\mathrm{PPC} \approx 2.11$--$2.19\sigma$, to be compared with the SPL-only value of $2.56\sigma$ obtained above.
As stressed above there is a basically irreducible $2.0\sigma$ tension below which one cannot go, corresponding to the tension between the KM3NeT event observation and the non-observation of neutrino events by the other experiments at the same energies (i.e.~within the ${\cal{I}}_{\text{in}}^{KM3}$ range).
This means that the KM3NeT event must be, in any case, interpreted as an overfluctuation at this minimum level of tension.
Thus, one finds that adding a Phenu component allows one to bring the tension close to the minimum achievable, i.e.~within $0.1\sigma$ of saturating the $2.0\sigma$ irreducible tension.\footnote{In refs.~\cite{Li:2025tqf, Palmisano:2025abd}, the tension between the KM3NeT event and the non-observation of similar events at IceCube has been quantified under various source assumptions, finding that diffuse and point source origins yield tensions of order $3\sigma$, with the exception of the specific case of a transient point source (see also \cite{Farzan:2025ydi}) that would not have been active during most of the IceCube observation time.}
This is also illustrated in figure~\ref{fig:results}, by the fact that the Phenu bump of each of the spectra peaks at the level of the joint UHE $E^{-2}$ best-fit flux within the energy range ${\cal I}_{\text{in}}^{\text{KM3}}$ (blue cross).
A $\sim2.1\sigma$ tension with respect to a $2.56\sigma$ tension improves the probability of the scenario by a factor $\sim3.3$ compared to the SPL-only fit.

The $(\tau_P = 10^{12}~\mathrm{s},\,m_P = 8\times10^{11}~\mathrm{GeV})$ spectrum improves on the SPL fit the least ($z_\mathrm{PPC} = 2.35\sigma$).
This is due to the fact that this case implies a neutrino flux larger than observed at lower energies $\sim$~PeV.
For this lifetime (and shorter ones), the large number of scatterings also makes the simulation significantly slower, resulting in fewer statistics and the wiggles visible in figure~\ref{fig:results} for this curve.
Also, in this case it is clear that the assumption that the Phenu component is negligible
in the HESE range is not correct anymore.
This explains why in table~\ref{tab:phenu} we do not consider even shorter lifetimes.
For lifetimes above $10^{13}$~s, by contrast, the results converge: $z_\mathrm{PPC}$ varies by less than $0.02\sigma$ across nearly four decades in $\tau_P$, since the sharp spectral feature is no longer appreciably affected by the scatterings on the \cnb.

\begin{table}[t]
\centering
\renewcommand{\arraystretch}{1.2}
\begin{tabular}{lllllll}
\hline
$\tau_P$ [s] & $m_P$ [GeV] & $\hat\phi_0$ & $\hat\gamma$
& $\hat{f}_P$ & $f_P^\mathrm{max}$ &  $z_\mathrm{PPC}$ \\
\hline
$10^{12}$           & $8\times10^{11}$    & $2.12$ & $2.87$
& $5.1^{+3.8}_{-4.7}\times10^{-6}$ & $1.1\times10^{-8}$   & $2.35$ \\
$3.16\times10^{12}$ & $5.33\times10^{11}$ & $2.12$ & $2.89$
& $1.5^{+2.0}_{-1.3}\times10^{-7}$ & $5.0\times10^{-10}$ & $2.19$ \\
$10^{13}$           & $4\times10^{11}$    & $2.12$ & $2.89$
& $1.3^{+2.5}_{-1.0}\times10^{-8}$ & $2.0\times10^{-10}$  & $2.12$ \\
$3.16\times10^{13}$ & $2.25\times10^{11}$ & $2.12$ & $2.89$
& $3.3^{+6.7}_{-2.7}\times10^{-9}$ & $2.6\times10^{-10}$  & $2.11$ \\
$10^{14}$           & $1.25\times10^{11}$ & $2.12$ & $2.89$
& $1.4^{+2.8}_{-1.1}\times10^{-9}$ & $5.5\times10^{-10}$ &  $2.12$ \\
$3.16\times10^{14}$ & $5\times10^{10}$    & $2.12$ & $2.89$
& $6.3^{+13}_{-5.1}\times10^{-10}$ & $1.7\times10^{-9}$ &  $2.12$ \\
$10^{15}$           & $2\times10^{10}$    & $2.12$ & $2.89$
& $2.9^{+5.8}_{-2.3}\times10^{-10}$ & $5.2\times10^{-9}$ & $2.12$ \\
$10^{16}$           & $4\times10^{9}$     & $2.12$ & $2.89$
& $5.9^{+12}_{-4.8}\times10^{-11}$ & $5.1\times10^{-8}$  & $2.12$ \\
$10^{17}$           & $8.5\times10^{8}$   & $2.12$ & $2.89$
& $8.8^{+18}_{-7.1}\times10^{-12}$ & $5.0\times10^{-7}$  & $2.12$ \\
\hline
\end{tabular}

\caption{Results of the Phenu + SPL joint fit for 9 values of the source particle lifetime.
$\hat\phi_0$ is in units of $10^{-18}~\mathrm{GeV^{-1}cm^{-2}s^{-1}sr^{-1}}$.
$f_P^\mathrm{max}$ is the upper bound on $f_P$ from CMB constraints as described in section~\ref{sec:constraints}.}
\label{tab:phenu}
\end{table}

Note finally that at the best-fit Phenu parameters, the Phenu flux at the reconstructed energy of KM3-230213A exceeds the HESE power-law best fit by a factor of $\sim$40.
The Phenu component nonetheless remains a subdominant contribution to the total expected event rate: the IceCube-EHE and Auger non-observation limits $f_P$ to values for which the expected Phenu count in $\mathcal{I}^\mathrm{KM3}_\mathrm{in}$ is well below unity.
What the Phenu component achieves is not to account for the observed event on its own (no isotropic diffuse flux can, given the null results of the other experiments) but to reduce the overfluctuation needed to explain the KM3NeT event to the minimum level possible.

\section{Constraints from cosmology}
\label{sec:constraints}
The decay of the heavy relic produces high energy neutrinos which, via FSR and 
subsequent EM and hadronic cascades, convert a significant fraction of their energy 
into EM radiation and hadronic material. For instance, for $m_P \simeq 10^{11}~\text{GeV}$, 
about 50\% of the injected energy eventually ends up as EM material (including photons and charged leptons), see~\cite{Bianco:2025boy}. 
This EM energy injection is subject to several cosmological 
constraints, discussed in detail in ref.~\cite{GrimbaumYamamoto:2025nai} 
(see also~\cite{Hambye:2021moy, Poulin:2016anj, Chluba:2020oip, Li:2025clq, Bianco:2025boy}), which we briefly summarise here.

The dominant constraint for the lifetimes considered in this work 
($\tau_P \gtrsim 10^{12}$~s) comes from CMB anisotropies. We use the 
bounds derived in~\cite{Hambye:2021moy}, based on~\cite{Acharya:2019uba}, which 
give an upper bound on $f_P$ as a function of $\tau_P$ and $m_P$. CMB 
spectral distortions also provide a constraint in this scenario (see~\cite{Poulin:2016anj, Chluba:2020oip}) but, as 
can be seen from figure~2 of~\cite{Hambye:2021moy}, they become relevant 
only for shorter lifetimes $\tau_P \lesssim 10^{12}$~s.\footnote{Although proposed future CMB spectral distortion experiments could significantly strengthen these constraints, making them competitive with or even dominant over the anisotropy bound~\cite{Li:2025clq}.} The $\Delta 
N_{\rm eff}$ constraint is less stringent than the CMB anisotropy bound 
(for masses above $\sim 100$~GeV, see figure~2 
of~\cite{Hambye:2021moy} or figure~8 of~\cite{GrimbaumYamamoto:2025nai}), 
and is therefore not relevant for the heavy relics considered here. BBN 
photo- and hadrodisintegration constraints~\cite{Bianco:2025boy,Hambye:2021moy} 
are most relevant for shorter lifetimes and higher masses, and are 
subleading compared to the CMB anisotropy bound for all $(\tau_P, m_P)$ 
values in table~\ref{tab:phenu}, as discussed 
in~\cite{GrimbaumYamamoto:2025nai}. 

The upper bounds on $f_P$ from CMB anisotropies, at the $2\sigma$ level are reported as $f_P^\mathrm{max}$ in table~\ref{tab:phenu}.
It is strongest for decays occurring near recombination, reaching $f_P^\mathrm{max} \simeq 2\times10^{-10}$ around $\tau_P \sim 10^{13}$~s, and weakens by three orders of magnitude by $\tau_P = 10^{17}$~s, as the injected energy has progressively less effect on the anisotropies.

\begin{figure}
    \centering
    \includegraphics[width=0.5\linewidth]{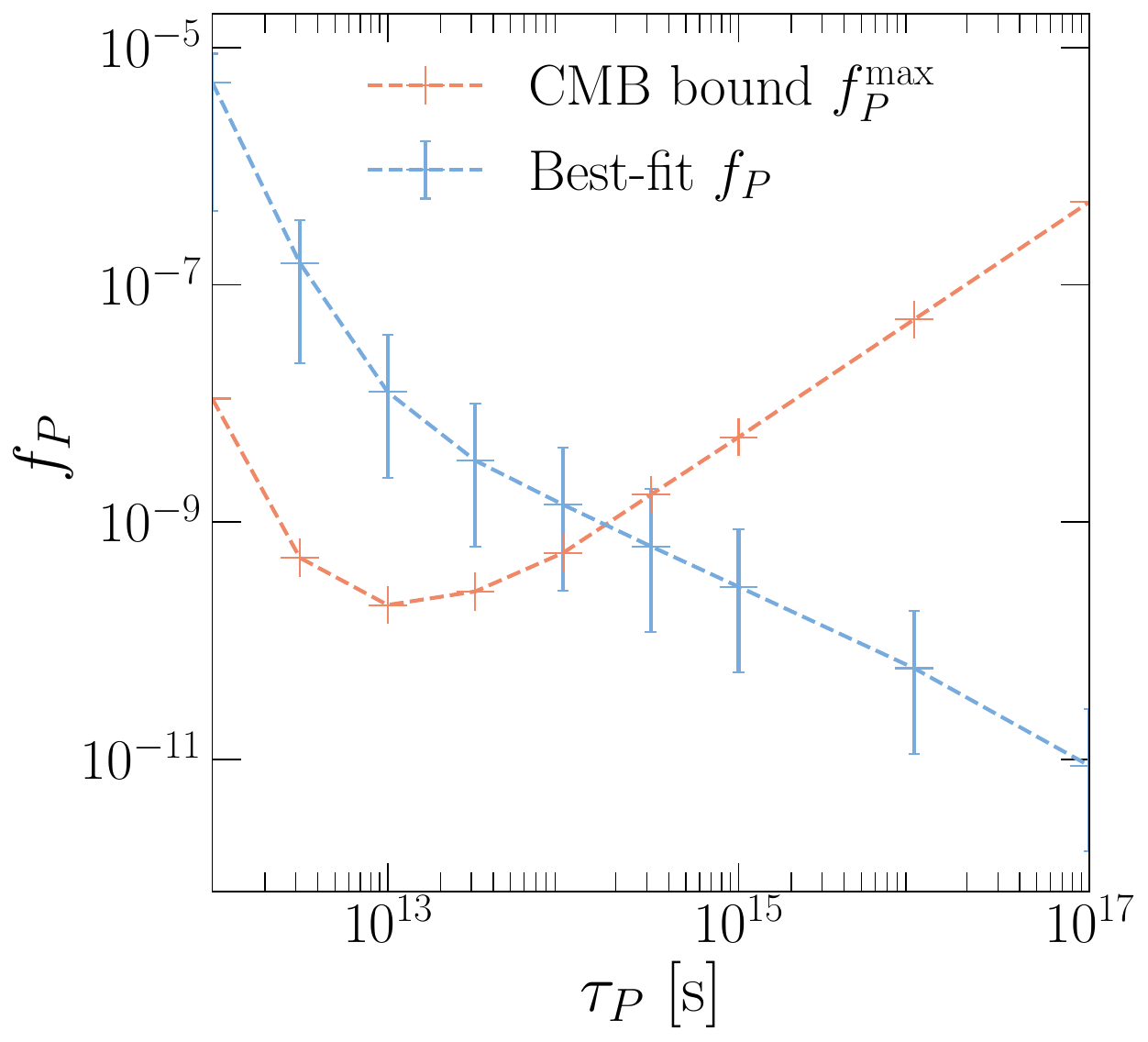}
    \caption{The Phenu + SPL best fit $f_P$ value from table~\ref{tab:phenu} (blue) against the CMB anisotropy upper bound (orange) as a function of the heavy relic lifetime. The CMB limit is strongest around recombination time $\sim10^{13}$s.}
    \label{fig:fp_cmb_fit}
\end{figure}

There is no a priori reason for this bound, which follows from EM energy injection at recombination, to fall anywhere near the abundance required to fit the KM3NeT event.
It is therefore notable that, in the lifetime range where the spectral fit is good, the two are of comparable size.
Lifetimes $\tau_P \gtrsim 10^{14}$~s are compatible with the CMB bound, while shorter ones are in tension with it as shown in figure~\ref{fig:fp_cmb_fit}.
The transition region, $\tau_P \simeq 10^{14}$~s, is the most promising one to leave an observable imprint in the CMB as it gives a best-fit $f_P$ around the central value of the bound (and compatible with it within the various uncertainties).
Next-generation CMB observatories would therefore either rule out this scenario for lifetimes around these values or observe an imprint of it in their data.

The possibility of a detection through CMB observatories is to be investigated as there are strong degeneracies with other BSM scenarios that can lead to energy deposition in the EM bath.
This is left for future work.
Since a good fit can also be obtained for a lifetime much larger than $t_\text{rec}$, where the anisotropy bound weakens considerably, late-time probes such as the 21-cm signal are also relevant although ref.~\cite{Mosbech:2026hlq} recently showed that 21-cm bounds are about an order of magnitude weaker than CMB anisotropy bounds for large relic masses (above 10 TeV).
Thus, CMB anisotropies remain the leading constraint on late-time injection of Phenus.

\section{\texorpdfstring{$\gamma$-ray constraints}{γ-ray constraints}}
\label{sec:checks}

It is also natural to ask whether gamma-ray observations place 
any additional constraint on $f_P$. To answer this, we perform an overly 
conservative back-of-the-envelope estimate. At the energies relevant for 
the Phenu spectra considered here, $\gamma$-rays interact efficiently with 
CMB photons via pair production and photon-photon scattering, and quickly 
cascade down to the so-called transparency window around $0.1$--$10$ 
GeV~\cite{Slatyer:2009yq}. From this point they mostly only redshift. 
To get a conservative upper bound on the production of $\gamma$-rays, one can thus assume that (unrealistically)
all of the energy of the decaying relics is 
instantaneously converted into $\gamma$-rays 
at the injection time, and that these $\gamma$-rays are deposited instantaneously in this 
transparency window. 
In this case the
electromagnetic energy density received today is
\begin{equation}
    u_\gamma^0 = f_P\,\Omega_{\rm DM}^0\,\rho_{\rm crit}^0\,,
    \label{eq:u_gamma}
\end{equation}
and the corresponding isotropic flux today is $\mathcal{F}_\gamma = u_\gamma^0/4\pi$.

For the Phenu spectra with lifetimes $\tau_P \lesssim t_{\rm rec}$, the 
injection occurs around recombination ($z_{\rm inj} \simeq 1100$), so that the 
transparency-window photons redshift to the $0.1$--$10$ MeV range today, 
precisely the band probed by COMPTEL~\cite{Weidenspointner:2000aq}.\footnote{Note 
that naively redshifting the secondary neutrinos ($\sim 10^{5}$ GeV) from the Phenu spectrum without cascading to the transparency window would 
place them in the Fermi-LAT range today.}
 We compare $\mathcal{F}_\gamma$ to the observed energy flux in the 
COMPTEL band, modelled as a power law $\mathrm{d}N/\mathrm{d}E = A 
(E/E_0)^{-2.4}$ with normalisation $A=(1.05\pm0.2)\times10^{-1}$ photons 
GeV$^{-1}$cm$^{-2}$s$^{-1}$sr$^{-1}$ at $E_0 = 5$ MeV~\cite{Kappadath:1998}, 
integrated over $[0.8\ \text{MeV},\ 30\ \text{MeV}]$. Equating 
$\mathcal{F}_\gamma = \mathcal{F}_{\rm COMPTEL}$ we find
\begin{equation}
    f_P \simeq \frac{4\pi\,\mathcal{F}_{\rm COMPTEL}}
    {\Omega_{\rm DM}^0\,\rho_{\rm crit}^0} \sim 1.5\times10^{-6}\,.
    \label{eq:fP_comptel}
\end{equation}

For spectra with longer lifetimes, e.g.\ $\tau_P \sim 10^{15}$ s, injection might 
occur as late as $z_{\rm inj} \sim 60$, so that photons sitting at the transparency window redshift 
to roughly $2-200$ MeV today, falling in the Fermi-LAT energy range. We compare 
$\mathcal{F}_\gamma$ to the energy flux of the isotropic gamma-ray background (IGRB) as measured by 
Fermi-LAT~\cite{Fermi-LAT:2014ryh}, with
\begin{equation}
    \frac{\mathrm{d}N}{\mathrm{d}E} = I_{100} 
    \left(\frac{E}{100\ \text{MeV}}\right)^{-\gamma}
    \exp\!\left(-\frac{E}{E_{\rm cut}}\right)\,,
\end{equation}
with parameters $I_{100} \simeq 10^{-4}\ \text{GeV}^{-1}\text{cm}^{-2}\text{s}^{-1}\text{sr}^{-1}$, 
$\gamma = 2.32$, $E_{\rm cut} = 279\ \text{GeV}$, integrated over 
$[2\ \text{MeV},\ 200\ \text{MeV}]$. Equating $\mathcal{F}_\gamma = 
\mathcal{F}_{\rm IGRB}$ we find
\begin{equation}
    f_P \simeq \frac{4\pi\,\mathcal{F}_{\rm IGRB}}
    {\Omega_{\rm DM}^0\,\rho_{\rm crit}^0} \sim 4\times10^{-7}\,.
    \label{eq:fP_fermi}
\end{equation}

These values of $f_P$ that 
would be necessary to produce an observable $\gamma$-ray flux by COMPTEL or Fermi-LAT, assuming in a conservative way that all the energy of the neutrinos would be converted into $\gamma$-rays, are three orders of magnitude larger than the ones that come out of the fits above, table~\ref{tab:phenu} and than the CMB upper bounds. 

\section{Summary}
\label{sec:summary}

We have analysed whether the ultrahigh energy neutrino observed recently by the KM3NeT collaboration could be a primordial high energy neutrino,
considering the example of a heavy relic decay into 2 neutrinos.
As long as the lifetime of the source particle is larger than a few times $10^{12}$ seconds it turns out that the Phenu spectrum in the detector displays a sharp spectral feature that could peak right at the KM3NeT event energy. This is an ideal situation to account for the KM3NeT event observation.\footnote{A spectral feature around 30~TeV has recently been observed in the diffuse astrophysical neutrino background measured by the MESE sample of IceCube~\cite{IceCube:2025dlr, IceCube:2025tgp}, favouring an SPL + bump scenario over a pure SPL. 
In a similar spirit to this work, it would be worthwhile to perform a dedicated analysis to see if such a bump could be due to a Phenu component.}
The determination of the shape of this peak required taking into account the effect of final state radiation, as well as of scatterings of these neutrinos with \cnb neutrinos all the way from production in the early Universe to Earth today. The latter effect required the development of a dedicated code. All in all, we find that the adjunction of a Phenu component to a power law (that is necessary to account for the observation of lower energy neutrinos around PeV by IceCube), improves the fit to about the best level one could aim for. There is, indeed, an irreducible experimental tension coming from the non-observation of neutrino events by the IceCube and Auger observatories at the KM3NeT energies, below which we cannot go. 
In practice it means that with respect to a power law case we improve the fit from 2.56$\sigma$ to about 2.1$\sigma$.
This does not constitute a huge improvement, but still this improves by a factor $\sim 3.3$ the probability of observing a neutrino event like the KM3NeT one.

One advantage of this scenario is that it does not lead to an associated flux of $\gamma$-rays that should already have been observed. 
Finally, for lifetimes around $10^{14}$~s, the source particle abundance that the best fit of the scenario requires is of the same order as the upper bound on it from CMB data.
This implies that, in that range and for larger lifetimes, the CMB would not have to be observed with a much greater sensitivity than today to either see an imprint of this scenario or rule it out.
Such an imprint would be of great importance as it would open a new window of observation on the Early Universe.

Note also importantly that the results obtained in this work for a heavy relic decay into 2 neutrinos, essentially remain unchanged for more general decays, for instance for a 2-body decay into a neutrino and any other SM particle, or for a 3-body decay with at least one neutrino in the final state. 
In the former (latter) case, the induced sharp spectral feature is the same (essentially the same) and at energies of order the KM3NeT event one, the $\gamma$-ray and CMB constraints are hardly changed. Thus, a Phenu explanation of the KM3NeT event could be found in a large variety of scenarios.
Last but not least, note that beyond the KM3NeT event, any observation of an isotropic distorted flux, if due to a flux of Phenus, would, through the application of the code we developed, provide indirect evidence for the C$\nu$B, and allow one to probe it.

\bmhead{Acknowledgements}
We thank S. Bianco, J. Frerick, L. Lopez Honorez, M. Hufnagel, K. Schmidt-Hoberg, and especially P. A. Sevle Myhr for useful discussions.
We also thank M. Ovchynnikov for pointing out a missing factor in the normalisation of the spectrum.
ULB report number: ULB-TH/26-07.
This work is supported by the Belgian IISN convention 4.4503.15 and by the Brussels laboratory of the Universe - BLU-ULB.
The work of NGY was supported by the Communaut\'e fran\c{c}aise de Belgique through a FRIA doctoral grant, and is currently supported by a Fondation Wiener-Anspach postdoctoral fellowship.

\appendix

\section{Data samples and statistical method}\label{app:statistics}

\begin{figure}
    \centering
    \includegraphics[width=0.5\linewidth]{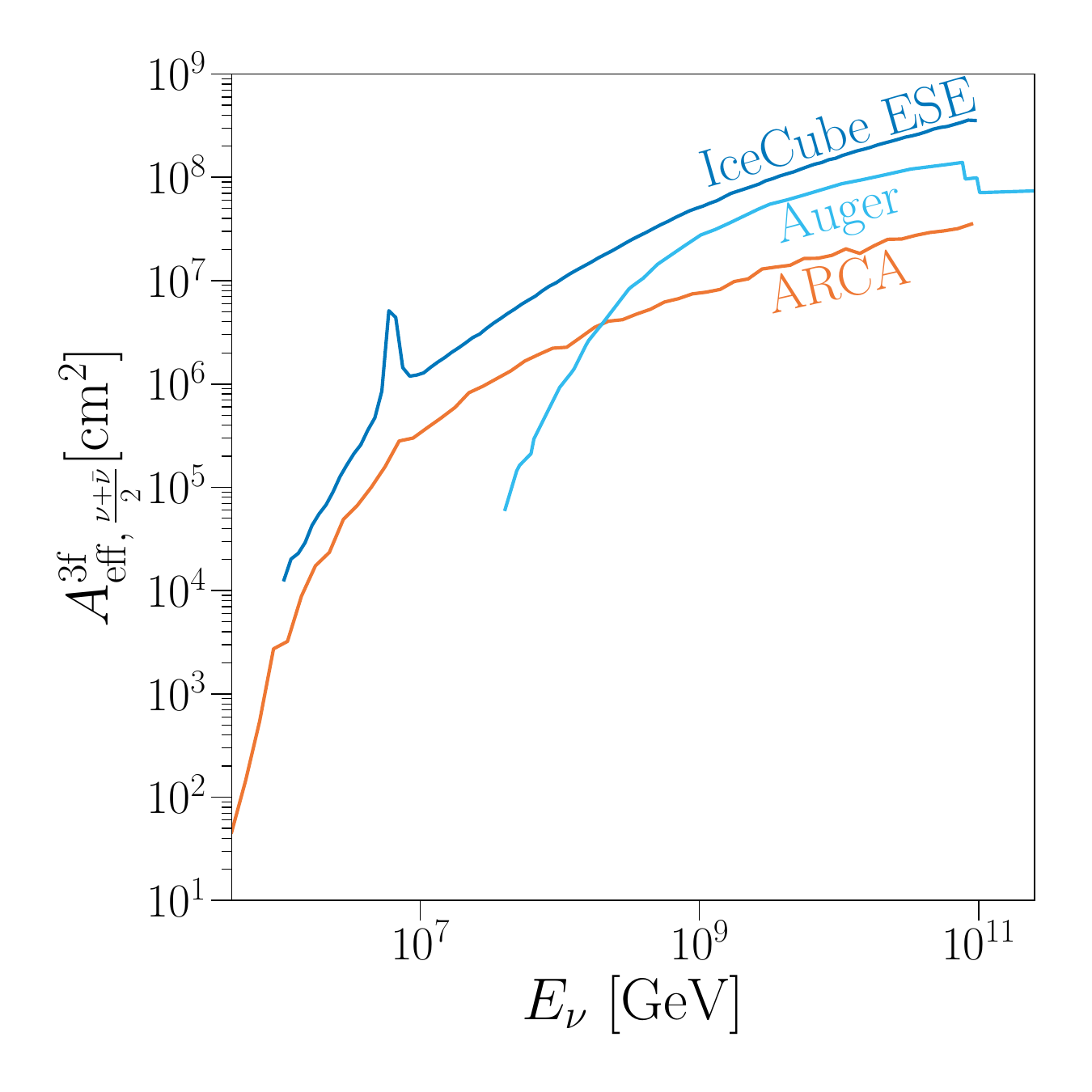}
    \caption{The sky averaged, all flavour, neutrino-antineutrino averaged effective area of KM3NeT/ARCA~\cite{KM3NeT:2025ccp}, IceCube EHE~\cite{IceCubeCollaborationSS:2025jbi} and Auger~\cite{PierreAuger:2019azx} used in the present analysis.}
    \label{fig:eff_area}
\end{figure}

The three UHE datasets that we consider, corresponding to the energy interval given in eq.~\eqref{eq:intervals}, are the ones of refs.~\cite{KM3NeT:2025npi,IceCubeCollaborationSS:2025jbi,PierreAuger:2019azx}. The corresponding exposures
$\mathcal{E}^d(E) = 4\pi \times T_d \times A^d_\mathrm{eff}(E)$
are constructed from the sky- and $\nu/\bar\nu$-averaged all-flavour effective areas $A_{\mathrm{eff}, \frac{\nu+\bar\nu}{2}}^{d,\text{3f}}$ (for a given dataset $d$) shown in figure~\ref{fig:eff_area}. 

For KM3NeT/ARCA we use the effective area from the public data release of
ref.~\cite{KM3NeT:2025npi} with $T_\mathrm{KM3} = 335$~days. For
IceCube-EHE we use the effective area and 12.6-year livetime ($T_\text{IC} = 4605$ days) from the
public data release of ref.~\cite{IceCubeCollaborationSS:2025jbi}; this differs from
ref.~\cite{KM3NeT:2025ccp}, which relied on the 9-year
analysis~\cite{IceCube:2018fhm}.
For the Pierre Auger Observatory, the effective area is computed by
summing contributions from the Earth-Skimming (ES,
$90$--$95^\circ$), Downward-Going High-angle (DGH,
$75$--$90^\circ$), and Downward-Going Low-angle (DGL,
$60$--$75^\circ$) channels from ref.~\cite{PierreAuger:2019azx},
each integrated over its zenith band via
$2\pi\int\mathrm{d}\theta\,\sin\theta\,A_\mathrm{eff}(E,\theta)$. The
ES channel is sensitive to the $\nu_\tau$ charged current (CC) channel only; DGH provides separate
effective areas for neutral current (all flavours), $\nu_e$, $\nu_\mu$, and
$\nu_\tau$ CC channels; DGL is sensitive to the $\nu_e$ CC channel only. 
Assuming a 1:1:1 flavour ratio, contributions are summed to a single
all-flavour effective area that is divided by $4\pi$ to account for the sky averaging. All three datasets provide the effective areas as averaged over neutrinos and antineutrinos. The Auger exposure uses the 18-year livetime of
ref.~\cite{PierreAuger:2023pjg}. 

In the following, we follow the statistical method defined in the appendix of ref.~\cite{KM3NeT:2025ccp} which we summarise here. The number of expected events given an exposure $\mathcal{E}^d(E)$ and a flux $\Phi(\Theta, E)$ is:
\begin{equation}
\label{eq:exp_event}
    \mu^d_\text{exp} (\Theta, \mathcal{I}_E) = \int_{\mathcal{I}_E}\mathcal{E}^d(E)\Phi(\Theta, E) dE \,,
\end{equation}
where $\Theta$ are the parameters of the flux model $\Phi$ and the energy intervals $\mathcal{I}_E$ are given in eq.~\eqref{eq:intervals}.
For the Auger and IceCube EHE datasets, we have that the number of reported events over $\mathcal{I}^\text{Auger/IC}$ are $N^\text{Auger/IC}_\text{obs} = 0$, while for the KM3NeT one we have $N^\text{KM3, in}_\text{obs} = 1$ and $N^\text{KM3, out}_\text{obs} = 0$. The contributions from the background for these UHE sets are neglected.

The constraints on the neutrino flux parameters $\Theta$ are obtained using a simple Poisson counting statistics, so that the likelihood is given by:
\begin{equation}
    \left\{\begin{aligned}
        \mathcal{L}_{d\text{, in}}(\Theta) &\equiv \mathrm{Poisson}\left(N^{d \text{, in}}_\text{obs}; \mu^d_\text{exp}\left[\Theta;\mathcal{I}_\text{in}(\Theta)\right]\right)\,, \\
        \mathcal{L}_{d\text{, out}}(\Theta) &\equiv \mathrm{Poisson}\left(N^{d \text{, out}}_\text{obs}; \mu^d_\text{exp}\left[\Theta;\mathcal{I}_\text{out}(\Theta)\right]\right)\,, \\
        \mathcal{L}_d(\Theta) &\equiv  \mathcal{L}_{d\text{, in}}(\Theta) \times \mathcal{L}_{d\text{, out}}(\Theta)\,,
    \end{aligned}\right.
\end{equation}
where $\text{Poisson}(N;\lambda) = e^{-\lambda}\lambda^N/N!$. Thus, both the Auger and IceCube datasets $\mathcal{L}_{d\text{, in}}(\Theta)$ and $\mathcal{L}_{d\text{, out}}(\Theta)$ have the same shape, with $N=0$, and we can directly compute $\mathcal{L}_\text{Auger/IC}$ over the whole interval $\mathcal{I}^\text{Auger/IC}$ with the observed number of events $N^\text{Auger/IC}_\text{obs} = 0$.

Note that the posteriors are evaluated on a regular grid in $(\phi_0, \gamma, \log_{10} f_P)$ throughout the analysis.
We checked that the quoted best-fit values are stable against halving the grid spacing.

Finally, from eq.~\eqref{eq:exp_event} we can express the product of the flux fitted by our analysis with the effective area as the sum over each neutrino and antineutrino flavour as $\sum_{i\in S} A^i_\text{eff} \Phi^i$, where $S = \{\nu_{e,\mu,\tau},\bar\nu_{e,\mu,\tau}\}$ is the set of all neutrino and antineutrino flavour. To express this in terms of the all-flavour neutrino-antineutrino averaged effective area $A^{\text{3f}}_{\text{eff}, \frac{\nu+\bar\nu}{2}} \equiv \sum_{i\in S}\frac{A^i_\text{eff}}{2}$, we assume a 1:1:1 flavour composition of the total, i.e. the all-flavour neutrino-antineutrino summed, flux. The latter can be expressed as $\Phi^\text{3f}_{\nu+\bar\nu} = 6\Phi^i$ and $\Phi^\text{3f}_{\nu+\bar\nu} = 3 \Phi^\text{1f}_{\nu+\bar\nu}$, where $\Phi^\text{1f}_{\nu+\bar\nu}$ is the per-flavour neutrino flux. The sum of the effective area and neutrino flux over each flavour becomes
\begin{equation}
    \sum_{i\in S} A^i_\text{eff} \Phi^i = \frac{ \Phi^\text{3f}_{\nu+\bar\nu}}{6}\times 2 A^{\text{3f}}_{\text{eff}, \frac{\nu+\bar\nu}{2}} = \Phi^\text{1f}_{\nu+\bar\nu}A^{\text{3f}}_{\text{eff}, \frac{\nu+\bar\nu}{2}}\,,
\end{equation}
such that the flux fitted using eq.~\eqref{eq:exp_event} is the per-flavour neutrino-antineutrino summed flux, given the all flavour neutrino-antineutrino averaged effective area.

The public HESE 7.5-year data sample we use is taken from~\cite{IceCube:2020wum}. Rather than re-implementing the HESE likelihood, we use the available SPL likelihood scan, provided as $\mathrm{TS}^\mathrm{(HESE)}_\mathrm{IC} =
-\log\mathcal{L}^\mathrm{(HESE)}_\mathrm{IC}$ as a function of the
all-flavour normalisation $\Phi_\mathrm{astro} \equiv \Phi^\text{3f}_{\nu+\bar\nu}(100\text{ TeV}) = 3\Phi^\text{1f}_{\nu+\bar\nu} (\text{100 TeV})$ (in units of
$10^{-18}~\mathrm{GeV^{-1}cm^{-2}s^{-1}sr^{-1}}$ at 100~TeV) and
$\gamma_\text{astro} \equiv \gamma$, converted to per-flavour via $\phi_0 =
\Phi_\mathrm{astro}/3$. The scan is tabulated for $\gamma_\text{astro} \in [1.5, 3.5]$, which is what fixes the range of the uniform prior on $\gamma$ used throughout.

\section{FSR at scattering}\label{app:fsr_scattering}

\begin{figure}
    \centering
    \includegraphics[width=0.5\linewidth]{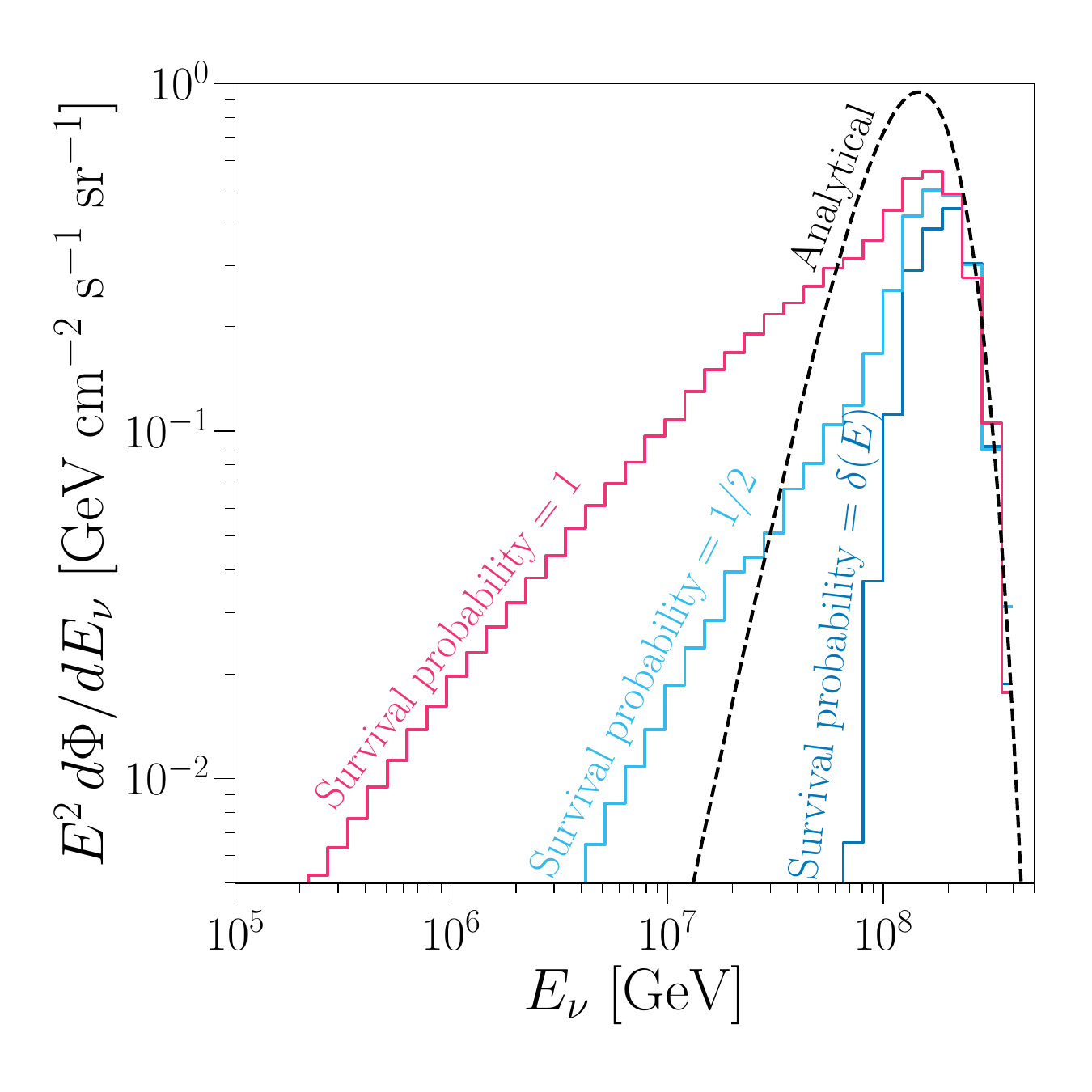}
    \caption{Impact of FSR at elastic scattering on the Phenu spectrum, for 
    $m_P = 5.33 \times 10^{11}$~GeV, $\tau_P = 3.2 \times 10^{12}$~s, 
    $f_P = 1$, retaining only elastic 
    scatterings in the Monte-Carlo simulation, for three assumptions on the survival probability 
    $P_\mathrm{surv}$ per scattering event: $P_\mathrm{surv} = 1$ (pink), 
    $1/2$ (cyan), and 
    $\delta(E_\nu)$ (blue). The analytical spectrum without scatterings or 
    FSR is shown as a dashed black curve for reference.}
    \label{fig:elastic_fsr}
\end{figure}

An effect we did not include in the code, and which is absent from all results above, is that of FSR associated with each scattering of the Phenu on the \cnb. 
Implementing this consistently requires the FSR spectrum for the process 
$\nu + \nu_\mathrm{bg} \to \nu + \nu_\mathrm{bg}$, which is not available 
from \texttt{HDMSpectra} (which models FSR in the context of dark matter decay and annihilation) 
and would require a dedicated treatment, e.g.\ using 
\texttt{PYTHIA}~\cite{Bierlich:2022pfr} at each scattering. This would appreciably complicate the code and slow down its efficiency.
However, we do not expect that these extra FSR effects of elastic scatterings could 
significantly affect the 
high energy peak of the Phenu spectrum, so that the fits performed above would hardly vary.

To estimate this effect, we compute the Phenu spectrum for $m_P = 5.33 \times 10^{11}$~GeV, 
$\tau_P = 3.2 \times 10^{12}$~s, $f_P = 1$, retaining only elastic 
scatterings and varying the probability $P_\mathrm{surv}(E_\nu)$ that a hard neutrino comes out of each elastic scattering despite the FSR.
Three cases are shown on figure~\ref{fig:elastic_fsr}: $P_\mathrm{surv} = 1$ (no FSR), 
$P_\mathrm{surv} = 1/2$ (motivated by the kinematics of a single $W/Z$ 
emission, after which the outgoing neutrino carries at most half the original 
energy) and $P_\mathrm{surv} = \delta(E_\nu)$,
where $\delta(E_\nu)$ 
is the fraction of neutrinos that do not undergo FSR, taken from 
\texttt{HDMSpectra} as a proxy, see figure~\ref{fig:FSR}.
We stress that, since \texttt{HDMSpectra} models a decay process 
rather than neutrino-neutrino scattering, the survival probability $\delta(E)$ shown 
is only illustrative. Also, from figure~\ref{fig:FSR}, it is clear that the high energy tail of the FSR spectrum would populate the peak of the Phenu spectrum too, so that keeping only the neutrinos that do not undergo FSR is an overly conservative assumption.
Note that the secondary neutrinos produced in FSR cascades, which 
would populate the spectrum at lower energies, are not included in 
figure~\ref{fig:elastic_fsr}. Their effect would be to widen the spectrum 
toward lower energies. 

The impact on the spectral peak is modest across most cases. This is 
expected: as discussed in ref.~\cite{GrimbaumYamamoto:2025nai}, the elastic 
scattering rate decreases at the high energy tail of the spectrum, due 
to the later emission time, so neutrinos near the peak undergo fewer 
scatterings and are less exposed to FSR during propagation. The only 
case showing a clear depletion of the peak is $P_\mathrm{surv} = 
\delta(E_\nu)$, which however represents an overly conservative estimate 
as it discards all elastic scatterings where FSR does occur, irrespective of their energy. Even in 
this case, the depletion remains below a factor of two. A realistic case is more likely to lie between the $P_\mathrm{surv}(E_\nu)=1$ and $P_\mathrm{surv}(E_\nu)=1/2$ lines.
Moreover, electroweak FSR only becomes relevant for $\sqrt{s} \gtrsim 100$~GeV, which, for a \cnb neutrino at the average energy of the Fermi-Dirac distribution, $\bar E_\text{BG} \simeq 3.15\,T_\nu$, requires a Phenu energy $E \gtrsim 4\times10^{12}~\text{GeV}\,[1100/(1+z)]$.
It therefore only affects the shortest lifetimes, $\tau_P \lesssim 10^{12}$~s, or otherwise scatterings on background neutrinos from the high-energy tail of the Fermi-Dirac distribution.
We 
conclude that neglecting FSR at elastic scattering introduces only a 
secondary uncertainty in the peak flux, and leave a proper implementation 
of this effect, as well as the treatment of EM and hadronic cascades initiated by inelastic scattering, for a future work.

\bibliography{biblio}

\end{document}